\documentclass[a4paper, 11pt]{article}
\usepackage[english]{babel}
\makeatletter

\usepackage{geometry,graphicx,color}
\usepackage{amsfonts, amsmath, amssymb, slashed,dsfont}
\usepackage{tensor}
\usepackage{hyperref}
\usepackage{epsfig}
\usepackage{pifont}
\usepackage{soul}
\usepackage{enumitem}
\usepackage{csquotes}
\usepackage{bm}
\usepackage{mathrsfs} 
\usepackage[hyperref,thmmarks,amsmath]{ntheorem}
\usepackage{amsmath,amssymb,slashed}
\usepackage{comment}
\usepackage{tikz}
\usetikzlibrary{positioning}
\usetikzlibrary{decorations.markings}
\usetikzlibrary{arrows.meta}
\usepackage{filecontents}
\usepackage{comment}
\numberwithin{equation}{section}
\newtheorem{th-imp}{Theorem}[section]

\DeclareMathOperator{\actr}{\triangleleft}
\DeclareMathOperator{\actl}{\triangleright}

\newcommand\bbone{\mathbb{I}}
\newcommand{\SU}{\mathrm{SU}(2)}
\newcommand\caZ{{\mathcal Z}}

\newcommand\algA{{\mathbb{A}}}
\newcommand\modM{{\mathbb{E}}}
\newcommand\algH{{\mathscr{H}}}
\newcommand\twiF{\mathscr{F}}

\newcommand\hR{{F}}
\newcommand\adrep{\mathrm{Ad}}

\newcommand\kS{{\mathfrak S}}

\DeclareMathOperator{\tr}{Tr} 
\newcommand{\sign}{\mathrm{sign}}
\DeclareMathOperator{\End}{\mathrm{End}}

\DeclareMathOperator{\Aut}{\mathrm{Aut}}

\newcommand\Int{\mathrm{Int}}
\newcommand\Out{\mathrm{Out}}
\newcommand\kX{X}
\newcommand\kY{Y}

\newcommand{\HS}{\mathrm{HS}}
\newcommand{\NN}{\mathbb{N}}
\newcommand{\CC}{\mathbb{C}}
\newcommand{\dd}{\mathrm{d}}

\newcommand\id{\mathrm{id}}

\newcommand\sstar{\text{\ding{86}}}

\newcommand{\institute}[1]{\newcommand{\@institute}{#1}}
\renewcommand{\maketitle}{
\vspace*{0.5\baselineskip}
{
\center\LARGE\noindent\@title\par
}%
\vspace{1.5\baselineskip}
{
\center\normalsize\noindent\ignorespaces\@author\par
}%
\vspace{0.5\baselineskip}
{
\center\normalsize\ignorespaces\@institute\par
}%
\vspace{2\baselineskip}
}%

\usepackage{siunitx}
\usepackage{xcolor}
\usepackage{booktabs,colortbl, array}
\usepackage{pgfplotstable}
\usepackage{pbox}
\definecolor{rulecolor}{RGB}{0,71,171}
\definecolor{tableheadcolor}{gray}{0.92}
\newcommand{\topline}{ %
        \arrayrulecolor{rulecolor}\specialrule{0.1em}{\abovetopsep}{0pt}%
        \arrayrulecolor{tableheadcolor}\specialrule{\belowrulesep}{0pt}{0pt}%
        \arrayrulecolor{rulecolor}}
\newcommand{\midtopline}{ %
        \arrayrulecolor{tableheadcolor}\specialrule{\aboverulesep}{0pt}{0pt}%
        \arrayrulecolor{rulecolor}\specialrule{\lightrulewidth}{0pt}{0pt}%
        \arrayrulecolor{white}\specialrule{\belowrulesep}{0pt}{0pt}%
        \arrayrulecolor{rulecolor}}
\newcommand{\bottomline}{ %
        \arrayrulecolor{white}\specialrule{\aboverulesep}{0pt}{0pt}%
        \arrayrulecolor{rulecolor} %
        \specialrule{\heavyrulewidth}{0pt}{\belowbottomsep}}%

\pgfplotstableset{normal/.style ={%
        header=true,
        string type,
        column type=l,
        every odd row/.style={
            before row=
        },
        every head row/.style={
            before row={\topline\rowcolor{tableheadcolor}},
            after row={\midtopline}
        },
        every last row/.style={
            after row=\bottomline
        },
        col sep=&,
        row sep=\\
    }
}

\begin{document}

\title{Noncommutative Gauge Theories: \\Yang-Mills extensions and beyond - An overview}
\author{Jean-Christophe Wallet}
\institute{%

\textit{IJCLab, Universit\'e Paris-Saclay, CNRS/IN2P3, 91405 Orsay, France}

\bigskip
e-mail:  

\href{mailto:jc.wallet.ups@gmail.com}{\texttt{jc.wallet.ups@gmail.com}}
}%

\maketitle

\begin{abstract} 
The status of several representative gauge theories on various quantum space-times, mainly focusing on Yang-Mills type extensions together with a few matrix model formulations is overviewed. The common building blocks are derivation based differential calculus possibly twisted and noncommutative analog of the Koszul connection. The star-products related to the quantum space-times are obtained from a combination of harmonic analysis of group algebras combined with Weyl quantization. The remaining problems inherent to gauge theories on Moyal spaces in their two different formulations are outlined. A family of gauge invariant matrix models on $\mathbb{R}^3_\lambda$, a deformation of $\mathbb{R}^3$ is presented among which a solvable model. The characterization of 11 new quantum Minkowski space-times through their $*$-algebras is given. A gauge theory of Yang-Mills type is constructed on one recently explored of these space-times and compared to its counterpart built on the popular $\kappa$-Minkowski.
\end{abstract}
\vskip 1 true cm
{\it{Mathematics Subject Classification}}: 81T75, 81T13, 81T32, 46L87, 17B37
\vfill\eject

\section{Introduction} \label{section1}
It is believed or expected that noncommutative structures \cite{connes1} may well emerge near the Planck scale as soon as Quantum Gravity effects \cite{zerevue}, \cite{whitepaper} would become non negligible so that the common notion of manifold should have to be traded for a quantum space-time, at least at some effective regime. Quantum space-times (aka noncommutative space-times) can be conveniently described using the concepts of noncommutative geometry \cite{connes1}. They have been the subject of a huge literature for more than 3 decades, ranking from mathematical developments to constructions of field theories on these quantum space-times, usually called noncommutative field theories. For a review on early works, see \cite{dnsw-rev}. It has been soon realized that their renormalization was often hard (or even not possible) to perform, due to additional difficulties, among which, apart from the problems linked to non locality, being the UV/IR mixing, which was however neutralized in some cases \cite{gw-a}, \cite{gw-b}.\\

Gauge theories on quantum space-times have appeared soon after their scalar counterpart, showing as it can be expected all the difficulties inherent to the noncommutative field theories, in particular for the renormalization, supplemented by the additional stringent constraints linked to gauge invariance. For a review on past mathematical and physical developments, see \cite{physrep}. In this note, I will overview the present status of some gauge theories on various quantum space-times, mainly focusing on Yang-Mills type extensions plus a few matrix model formulations. Their common features is to be associated to a derivation-based differential calculus \cite{mdv} while their underlying notion of connection is the usual noncommutative analog of the Koszul connection on a (right-)module \cite{KOSZUL-T}, \cite{mdv}.\\

The paper is organized as follows. The section \ref{section2} collects all the mathematical material which underlies the gauge theories to be considered. A construction of the star-products used in the paper whose expressions are especially convenient in field theory and based on standard properties of harmonic analysis of group algebras combined with Weyl quantization is presented. Properties of derivation-based differential calculus together with the noncommutative analog of the Koszul connection used in the paper are recalled. The section \ref{section3} deals with gauge theories on Moyal spaces $\mathbb{R}^{2n}_\theta$ and $\mathbb{R}^{3}_\lambda$. The UV/IR mixing problem occurring in the Yang-Mills version on $\mathbb{R}^{4}_\theta$ and the various attempts to overcome it are discussed. The vacuum problem appearing in the matrix model version is presented.  The gauge invariant action is somehow similar to the action of a family of type IIB matrix models \cite{matrix1-a}, \cite{matrix1-b}. Gauge invariant matrix models on $\mathbb{R}^{3}_\lambda$ can accomodate a harmonic term as in the Grosse-Wulkenhaar model \cite{gw-a} and are finite to all orders. A particular solvable model is also presented.
In the section \ref{section4}, the recent construction of the star-products and involution characterizing the $*$-algebras for 11 new quantum Minkowski space-times, having "noncommutativity of Lie-algebra type" is summarized. These quantum space-times are related \cite{mercati-a}, \cite{mercati-b} in a particular sense to quantum deformations of the Poincar\'e symmetry related to the Poisson structures supported by the Poincar\'e group, classified a long time ago \cite{zak}. One gauge theory of Yang-mills type is constructed on $\rho$-Minkowski \cite{gauge-rho} recently explored \cite{marij-a}, \cite{marij-b} belonging to these new quantum space-times and compared to its counterpart built on the popular $\kappa$-Minkowski \cite{luk2} for which the gauge invariance fixes uniquely the number of dimensions \cite{MW2020a}. The section \ref{section5} discusses the results.
 \section{Algebraic and technical aspects}\label{section2}
\subsection{Star-products and quantum symmetries}\label{section21}
A quantum (i.e. noncommutative) space-time can be described as an associative $*$-algebra of (suitably behaving) functions, says $\mathbb{A}$. Most of the related associative products, namely the star-products, which are used in the physics literature belong to the category of formal deformations of the commutative point-wise product of functions and can be written as a formal expansion in some deformation parameter, often identified to a huge mass scale, with functions as coefficients of the expansion.\\

The star-products can be constructed from different ways. One popular way is based on the twist deformation \cite {Majid_1995} of the commutative point-wise product of functions of an algebra, says $\mathcal{A}$. This construction uses intensively the notion of Hopf algebra (see e.g. \cite {Klimyk_1997} for relevant material) which in some sense models the action of the "quantum symmetries" of the quantum 
space-time. Recall that a Hopf algebra $\algH$ can be deformed into another Hopf algebra $\algH^\twiF$, using a twist. It is an invertible element $\twiF \in \algH \otimes \algH$ with
\begin{subequations}
\begin{align}
    (\twiF \otimes 1) (\Delta \otimes \id) (\twiF)
    &= (1 \otimes \twiF) (\id \otimes \Delta) (\twiF), &
    \text{($2$-cocycle condition)}
    \label{2-cocycle} \\
    (\id \otimes \varepsilon) (\twiF)
    &= (\varepsilon \otimes \id) (\twiF)
    = 1. &
    \text{(normalisation)}
    \label{normalisation}
\end{align}
    \label{eq:twist_def}
\end{subequations}
Then, it can be shown that $\algH^\twiF= (\algH, \mu, \eta, \Delta^\twiF, \varepsilon, S^\twiF)$ is also a Hopf algebra, where
\begin{align}
    \Delta^\twiF(h)
    &= \twiF \Delta(h) \twiF^{-1}, &
    S^\twiF(h)
    &= \chi S(h) \chi^{-1}\label{quant-twist},
\end{align}
with $\chi = \twiF_1 S(\twiF_2)$, upon writing $\twiF = \twiF_1 \otimes \twiF_2$ and $h \in \algH$. \\
Now the action of the Hopf algebra $\algH$ on the space-time algebra $\mathcal{A}$ is modeled by turning $(\mathcal{A}, \cdot, \bbone)$ into an $\algH$-module algebra\footnote{
The following definition is written for a left action $\actl$, however, it could equivalently be written for a right action $\actr$.
}, \textit{i.e.}\ it must satisfy
\begin{subequations}
\begin{align}
    (hg) \actl a &= h \actl (g \actl a), &
    1 \actl a &= a,
    \label{hmodule1}\\
    h \actl (a \cdot b) &= (h_{(1)} \actl a) \cdot (h_{(2)} \actl b), &
    h \actl \bbone &= \varepsilon(h) \bbone,
    \label{hmodule2}
\end{align}
    \label{eq:hopf_module_alg}
\end{subequations}
where $\actl: \algH \otimes \mathcal{A} \to \mathcal{A}$ is the action of $\algH$ on $\mathcal{A}$, $a,b \in \mathcal{A}$ and $h, g \in \algH$. Then, one can show that the twisted Hopf algebra $\algH^\twiF$ acts on an algebra $\algA$ defined by $(\mathcal{A}, \star, \bbone)$, hence trading the point-wise product $\cdot$ for a new associative (but noncommutative) product defined by
\begin{align}
    a \star b
    &= \cdot \circ \twiF^{-1} \actl (a \otimes b)
    = (\twiF^{-1}_1 \actl a) \cdot (\twiF^{-1}_2 \actl b).
    \label{star-twist}
\end{align}
One advantage of the twist approach is that it leads to the simultaneous characterization of both the quantum space and its "quantum symmetries", hence the two objects in the pair $(\mathcal{H},\mathbb{A})$ are rigidly linked. However, the obtained star-product as expressed as an infinite expansion which stems directly from \eqref{star-twist} is hardly suitable to practical exploration of field theories.\\
An alternative but not necessarily equivalent construction is based on the use of basic properties of harmonic analysis of locally compact groups linked to the noncommutative algebras of coordinates (in the physicists language), in particular their related convolution algebras, combined with the Weyl quantization. This approach is in fact directly inherited from pioneering works of von Neumann and Weyl \cite{vonNeum},\cite{Weyl}.\\
In the following, the groups of interests have a semi-direct product structure of the form
\begin{equation}
    \mathcal{G}
    := H\ltimes_{{\phi}} \mathbb{R}^{n}
    \label{genesemidirect}
\end{equation}
$n\ge1$, where $\mathbb{R}^n$ is the additive group of real numbers, $H$ is a subgroup of $GL(n,\mathbb{R})$ and $\phi: H \to \mathrm{Aut}(\mathbb{R}^{d})$ defines the usual action of any matrix in $H\subset GL(n,\mathbb{R})$ on $\mathbb{R}^n$, namely
\begin{equation}
    \phi_a(x) = a x,
    \label{phiaction}
\end{equation}
for any $a\in H$, $x\in\mathbb{R}^n$. The structure of $\mathcal{G}$ is defined by
\begin{align}
    (a_1, x_1) (a_2, x_2)
    &= (a_1 a_2, x_1 + a_1 x_2), &
    \label{w1}\\
    (a,x)^{-1}
    &= (a^{-1}, -a^{-1} x), &
    \bbone_{\mathcal{G}}
    = (\bbone_H,0)
    \label{w2},
\end{align}
where $(a,x)$ denotes generically an element of $\mathcal{G}$. The related convolution product is
\begin{equation}
    (F\circ G)(s)
    = \int_{\mathcal{G}}\ d\mu(t) F(s t) G(t^{-1})
    \label{convol-gene}
\end{equation}
for any $F,G\in L^1(\mathcal{G})$, $s,t\in\mathcal{G}$, where $d\mu(t)$ is the left-invariant Haar measure, related to the right-invariant Haar measure, says $d\nu$, by $d\nu(s)=\Delta(s^{-1})d\mu(s)$ for any $s\in\mathcal{G}$, where the group homomorphism $\Delta:\mathcal{G}\to\mathbb{R}^+$ is the modular function, with 
$\Delta(s)=1$ for unimodular groups. Textbook properties of semi-direct 
product $\mathcal{G}$ \eqref{genesemidirect} yield the following expression for the Haar measure and modular function of $\mathcal{G}$ (in obvious notations)
\begin{equation}
    d\mu_\mathcal{G}((a,x))
    = d\mu_{\mathbb{R}^n}(x)\ d\mu_H(a)\ |\det(a)|^{-1}
    \label{measure},
\end{equation} 
\begin{equation}
    \Delta_\mathcal{G}((a,x))
    = \Delta_{\mathbb{R}^n}(x)\ \Delta_H(a)\ |\det(a)|^{-1}
    \label{modul-function}
\end{equation}
for any $a\in H$, $x\in\mathbb{R}^n$ where $d\mu_{\mathbb{R}^n}(x)$ is the Lebesgue measure on $\mathbb{R}^n$, i.e. $d^nx$, and $d\mu_H$ and $\det(a)$ depend on the choice of $H$. The convolution algebra, denoted hereafter by $\mathbb{C}(\mathcal{G}):=(L^1(\mathcal{G}),\circ,^\sstar)$ is a $\star$-algebra thanks to the natural involution defined by
\begin{equation}
    F^\sstar(x)
    = {\overline{F}}(x^{-1}) \Delta_\mathcal{G}(x^{-1})
    \label{involution}
\end{equation}
for any $F\in L^1(\mathcal{G})$, $x\in\mathcal{G}$, where ${\overline{F}}$ is the complex conjugate of $F$. Given a unitary representation of $\mathcal{G}$ $\pi_U:\mathcal{G}\to\mathcal{B}({\mathcal{H}})$, the induced $\star$-representation of $\mathbb{C}(\mathcal{G})$ on $\mathcal{B}({\mathcal{H}})$, $\pi:\mathbb{C}(\mathcal{G})\to\mathcal{B}({\mathcal{H}})$, is given by $ \pi(F)
   = \int_\mathcal{G} d\mu_\mathcal{G}(x) F(x) \pi_U(x)$, for any $F\in\mathbb{C}(\mathcal{G})${\footnote{$F$ must be compactly supported. Note that $\pi_U$ must be strongly continuous, which is the case here.}} and is bounded and non-degenerate. Thus, one can write
\begin{align}
    \pi(F\circ G)
    = \pi(F)\pi(G), &&
    \pi(F)^\ddag
    = \pi(F^\sstar)
    \label{pi-morph}
\end{align}
with $\pi(F)^\ddag$ the adjoint operator of $\pi(F)$. Now, set $F=\mathcal{F}f$ and $G=\mathcal{F}g$, where $\mathcal{F}$ denotes the Fourier transform\footnote{
    Convention for the Fourier transform: $\mathcal{F}f(p) = \int \frac{d^dx}{(2\pi)^d}\ e^{- i p x} f(x)$ and $f(x) = \int d^dp\ e^{i p x} \mathcal{F}f(p)$.
}.thus assuming that {\it{the elements of $\mathbb{C}(\mathcal{G})$ are functions on a momentum space}}. This, combined with the Weyl quantization operator given by
\begin{equation}
    Q(f)
    = \pi(\mathcal{F}f)
    \label{weyl-operat}
\end{equation}
leads to $Q(f\star g)
    = Q(f) Q(g)$ and $
    (Q(f))^\ddag
    = Q(f^\dag)$, from which follows
\begin{align}
    f\star g
    = \mathcal{F}^{-1} (\mathcal{F}f \circ \mathcal{F}g), && 
    f^\dag
    = \mathcal{F}^{-1} (\mathcal{F}(f)^\sstar)
    \label{star-prodetinvol},
\end{align}
while the algebra of functions $\mathcal{F}^{-1}F$, $F\in\mathbb{C}(\mathcal{G})$ endowed with the star-product and involution \eqref{star-prodetinvol} is interpreted as the $*$-algebra of functions modeling the quantum space-time whose noncommutative algebras of coordinates is related to the group $\mathcal{G}$.

\subsection{Differential calculus, connection and curvature}\label{section22}

Most of the gauge theories presented in this note are described by using different versions of the derivation-based differential calculus, introduced and developed a long time ago. For mathematical details see e.g. \cite{mdv}. It appears to be well suited to formulate conveniently quantum field theories on quantum space-times. Other interesting noncommutative differential calculi have also been used in the physics literature, as those obtained from twist deformations of a classical differential calculus \cite{Majid_1999-a}, \cite{Majid_1999-b}. The rest of this subsection will list the useful elements relevant for the ensuing discussion.\\

One starts with $\mathrm{Der}(\algA) $, the linear space of the derivations of $\algA$, \textit{i.e.}\ the linear maps $\kX : \algA \rightarrow \algA$ satisfying the Leibniz rule for any $a,b\in \algA$:
\begin{equation}
     \kX(a \star b) 
    = \kX(a) \star b + a \star \kX(b).
\end{equation}
$\mathrm{Der}(\algA)$ is Lie algebra when equipped with the bracket $[\kX, \kY ](a) = \kX  (\kY (a)) - \kY (\kX (a))$ and a $\caZ(\algA)$-module for the action $(z \actl \kX) (a) = z \star \kX (a)$, for any $a\in\algA$, $z \in \caZ(\algA)$, $\kX, \kY \in \mathrm{Der}(\algA)$, where $\caZ(\algA)$ is the center of $\algA$.\\
The linear subspace $\Int(\algA) \subset \mathrm{Der}(\algA)$ involving derivations such that $ \adrep_a : b \mapsto [a,b],\ a \in \algA$ is called the space of inner derivations. $\Int(\algA)$ is a $\caZ(\algA)$-submodule. Derivations in $\mathrm{Der}(\algA)$ which are not inner, \textit{i.e.}\ $\Out(\algA) = \mathrm{Der}(\algA) / \Int(\algA)$, are called outer derivations.\\

The derivation-based differential calculus is defined from the space of $\caZ(\algA)$-multilinear antisymmetric maps $\omega: \mathrm{Der}(\algA)^n\to\algA$, $n\in\mathbb{N}$, denoted by ${\Omega}^n(\algA)$, with ${\Omega}^0(\algA)=\algA$. Set $ \Omega^\bullet(\algA) 
    = \bigoplus_{n \geqslant 0} \Omega^n(\algA).$
The product $\wedge: \Omega^\bullet(\algA) \to \Omega^\bullet(\algA)$ for any $\omega\in{\Omega}^p(\algA), \eta\in{\Omega}^q(\algA)$ is:
\begin{equation}
\begin{aligned}
    (\omega \wedge \eta) & (\kX_1, \dots, \kX_{p+q}) \\
    &:= \frac{1}{p!q!} \sum_{\sigma\in \kS_{p+q}} (-1)^{{\sign}(\sigma)}
    \omega(\kX_{\sigma(1)}, \dots, \kX_{\sigma(p)}) \star \eta(\kX_{\sigma(p+1)}, \dots, \kX_{\sigma(p+q)}),
    \label{form-product-basic}
\end{aligned}
\end{equation}
and the differential $\dd: \Omega^p(\algA) \to \Omega^{p+1}(\algA)$ is
\begin{align}
\begin{aligned}
    \dd \omega(\kX_1, \dots, \kX_{p+1}) 
    :=& \sum_{j = 1}^{p+1} (-1)^{j+1} \kX_j \big( \omega( \kX_1, \dots, \vee_j, \dots, \kX_{p+1}) \big) \\
    &+ \sum_{1 \leqslant j < k \leqslant p+1} (-1)^{j+k} \omega( [\kX_j, \kX_k], \dots, \vee_j, \dots, \vee_k, \dots, \kX_{p+1}),
    \label{differential-basic}
\end{aligned}
\end{align}
for any $\omega\in{\Omega}^p(\algA)$, in which $\vee_j$ denotes the omission of the element $X_j$. One has $\dd^2=0$.\\
The differential satisfies the following expected graded Leibniz rule
\begin{equation}
    \dd(\omega \wedge \eta)
    = \dd\omega \wedge \eta + (-1)^{|\omega|} \omega \wedge \dd\eta,
    \nonumber
\end{equation}
for any $\omega,\eta\in\Omega^\bullet(\algA)$ where $|\omega|$ denotes the degree of $\omega$. Then, one verifies that the triplet
\begin{equation}
    (\Omega^\bullet(\algA), \wedge, \dd)
    \nonumber
\end{equation}
is a (${\mathbb{N}}$-graded) differential algebra defining the derivation-based differential calculus.\\
At this stage, two comments are in order:\\
i) For physical consideration, it is often convenient to restrict the set of derivations to a Lie subalgebra of $\mathrm{Der}(\algA)$, thus working with a so-called restricted differential calculus, characterized by a trivial modification of the above scheme. Various restricted differential calculi will be used in the next sections.  \\
ii) The above scheme can be straightforwardly adapted to the case of twisted derivations, as it will be shown in section 4.\\

A noncommutative connection can be defined as a noncommutative analog of the Koszul connection on a vector bundle, i.e. a linear map $\nabla: \Gamma(\mathcal{E}) \to \Gamma(\mathcal{E} 
    \otimes T^*\mathcal{M})$ verifying the relation $\nabla(mf)=\nabla(m)f+m\otimes df$, for any $m\in\Gamma(\mathcal{E})$ the space of sections of a vector bundle $\mathcal{E}$ over a (smooth) manifold $\mathcal{M}$, $f\in C^\infty({\mathcal{M}})$ and $T^*\mathcal{M}$ the cotangent bundle. Recall that $\Gamma(\mathcal{E})$ is a module over $C^\infty({\mathcal{M}})$ with action being the point-wise product. From this, one get the equivalent description, more convenient for the formulation of field theories, as:
\begin{equation}
\nabla_X: \Gamma(\mathcal{E}) \to \Gamma(\mathcal{E}),\ \ \nabla_X(m f) = m X(f) + \nabla_X(m) f,\ \ \label{koszul-connect}
\end{equation}
for any $X\in \Gamma(\mathcal{M})$, the Lie algebra of vector fields.\\

A well-known noncommutative analog of the above gives rise to a noncommutative connection \cite{mdv} defined as a linear map $\nabla_\kX : \modM \to \modM$ 
where $\modM$ is a right module over $\mathbb{A}$, the associative algebra modeling the quantum space-time, satisfying
\begin{equation}
    \nabla_\kX (m \actr a) 
    = m \actr \kX(a) + \nabla_\kX (m) \actr a,\ \nabla_{z \star \kX + \kY}(m) 
    = \nabla_\kX(m) \actr z + \nabla_\kY(m)
    \label{nc-connexion}
\end{equation}
for any $\kX,\kY \in \mathrm{Der}(\algA)$, $a \in \algA$, $m \in \modM$, $z \in \caZ(\algA)$ the center of $\algA$, where $\actr: \modM \otimes \algA \to \modM$ is the right action of the $\mathbb{A}$-module $\mathbb{E}$. Note that this extends to $\nabla: \modM \to \modM \otimes_\algA \Omega^1(\algA)$, $\nabla(m \actr a)= \nabla(m) \actr a + m \otimes \dd a$ which can be further extended to a map $\nabla: \modM \to \modM \otimes_\algA \Omega^\bullet(\algA)$.\\
The curvature is defined as the following morphism of module 
\begin{equation}
    \hR(\kX, \kY) : \modM \to \modM,\ \ \hR(\kX, \kY) (m) 
    = [\nabla_\kX, \nabla_\kY ] (m) - \nabla_{[\kX, \kY]}(m)
    \label{eq:nc_curv_def},
\end{equation}
for any  $m\in\modM$, $\kX, \kY \in \mathrm{Der}(\algA)$.\\

Note that the above holds for any restricted set of derivations of $\mathrm{Der}(\algA)$ with a Lie algebra structure. In the following, any suitable Lie algebra of derivations will be denoted by $\mathcal{D}$. Otherwise stated, any derivation will be further assumed to be real, i.e. $X(a)^\dag=X(a^\dag)$ in 
obvious notations.\\

To deal with a noncommutative analog of a hermitian connection, it is convenient to equip $\modM$ with a hermitian structure $h:\mathbb{E}\otimes\mathbb{E}\to\mathbb{A}$ with $h(m_1,m_2)^\dag=h(m_1^\dag,m_2^\dag)$ and $h(m_1 \actr a_1,m_2 \actr a_2)=a_1^\dag h(m_1,m_2)a_2$ for any $m_1,m_2\in\mathbb{E}$, $a_1,a_2\in\mathbb{A}$. A hermitian connection is then defined by requiring that $ X \big( h(m_1, m_2) \big)
    = h \big(\nabla_X(m_1), m_2 \big) + h \big(m_1, \nabla_X(m_2) \big)$ holds true for any $X\in\mathcal{D}$.\\
    
The gauge transformation on a connection are defined as the group of automorphisms of the right module over $\mathbb{A}$, $\varphi \in \Aut(\modM)$, such that
$\nabla^\varphi_\kX 
    = \varphi^{-1} \circ \nabla_\kX \circ \varphi$ is still a connection. This gives rise to the gauge transformation of the curvature which is expressed as $ \hR^\varphi(X,Y)
    = \varphi^{-1} \circ \hR(X,Y) \circ \varphi$, for any $X,Y\in\mathcal{D}$.\\
It is convenient to further require that the gauge transformations are compatible with the hermitian structure which is expressed as $h(\varphi(m_1),\varphi(m_2))=h(m_1,m_2)$ for any $\varphi\in\Aut(\modM)$, $m_1,m_2\in\modM$, thus introducing a noncommutative analog of unitary gauge transformations.\\

In the following, the $\mathbb{A}$-module $\mathbb{E}$ will be assumed to be one copy of the algebra $\mathbb{A}$, i.e. $\modM \simeq \algA$, an assumption which covers most of the physics literature on noncommutative field theories. Besides, the hermitian structure will be assumed to be the canonical one, namely: $h(m_1,m_2)=m_1^\dag\star m_2$, the symbol $^\dag$ denoting the involution of $\mathbb{A}$. Then, it is straightforward to characterize the group of gauge transformations $\mathcal{U}(\modM\simeq\mathbb{A})$ as 
\begin{equation}
    \mathcal{U}(\algA) 
    = \{g \in \algA,\ g^\dag \star g = g \star g^\dag = \bbone\},
    \label{gauge-transf}
\end{equation}
where $\varphi(\bbone) = g \in \modM\simeq\mathbb{A}${\footnote{It will be assumed that $\mathbb{E}$ is equipped with a unit or an approximate unit. }}. Besides, the noncommutative {\it{hermitian}} connection and corresponding curvature are determined (in obvious notations) by
\begin{equation}
    \nabla_X(a)=X(a)+A_X\star a,\ \ A_X:=\nabla_X({\bbone}),\ \ A_X^\dag=-A_X\label{amu-def}
    \end{equation}
    \begin{equation}
        F(X,Y)=X(A_Y)-Y(A_X)+[A_X,A_Y]_\star
    \end{equation}
for any $a\in\mathbb{A}$, $X,Y\in\mathcal{D}$ where the symbol $\star$ denotes the associative product in $\mathbb{A}$. This leads to the following gauge transformations:
\begin{align}
    A^g_X
    = g^\dag \star A_X \star g + g^\dag \star X(g),\ \ \hR(X,Y)^g 
    = g^\dag \star \hR(X,Y) \star g,
    \label{gauge-practical}
    \end{align}
for any  $X, Y\in\mathrm{Der}(\algA)$ and any $g \in \mathcal{U}(\algA)$.\\

\section{Gauge theory models on Moyal spaces and deformed $\mathbb{R}^3$}\label{section3}

In this section, I will summarize and comment critically the main features of some prototypal gauge theory models on popular quantum spaces which have received a considerable interest in the two past decades. These  quantum spaces are the Moyal spaces of dimension $2n$, very roughly a " product of $n$ phase spaces" in view of the corresponding algebra of coordinates and a deformation of the 3-d Euclidean space, sometimes emerging in developments of Loop Quantum Gravity or Group Field Theory, which can be viewed as an infinite direct sum of fuzzy spheres. One of the main issues was the construction of a perturbatively renormalizable theory to all order which would have become an actual noncommutative analog of a usual Yang-Mills theory. This gave rise to two different approaches, depending on the type of field variable used, either being the noncommutative analog of the gauge potential as defined in \eqref{amu-def} leading to a mere extension of Yang-Mills theories or being a tensor form with covariant gauge transformations as a consequence of the occurrence of inner derivation in $\mathcal{D}$ as recalled below, leading to a description in term of matrix models.\\

The present conclusion is disappointing. Despite a huge amount of effort, all attempts to obtain a gauge theory model perturbatively renormalisable to all orders and with a suitable commutative limit failed so far. This comes either from a lack of method to overcome the UV/IR mixing, still present as in most of the noncommutative field theories or to overwhelming technical difficulties stemming from the complicated structure of the vacuum showing up in the matrix model description.

\subsection{Generalising Yang-Mills theory on Moyal spaces}\label{section31}

There is a huge available literature on the Moyal spaces \cite{graciavar1-a}, \cite{graciavar1-b}. Informally, the Moyal space, hereafter generically denoted by $\mathbb{R}^{2n}_\theta$ can be viewed as the universal enveloping algebra of
\begin{equation}
    [x^\mu, x^\nu]_\theta
    = i \Theta^{\mu\nu}\ \ , \ \Theta
    = \theta\ \mathrm{diag}(J,\dots, J), 
    J 
    = \begin{pmatrix} 0 & 1 \\ -1 & 0 \end{pmatrix}\label{alg-Moyal},
    \end{equation}
where $[x^\mu, x^\nu]_\theta: = x^\mu \star_\theta x^\nu - x^\nu \star_\theta x^\mu$
and $\star_\theta$ is the star-product recalled below. $\mathbb{R}^{2n}_\theta$ is usually described by an associative $*$-algebra $\mathbb{R}^{2n}_\theta=(\mathcal{M}(\mathbb{R}^{2n}),\star_\theta,\dag)$ where the involution $\dag$ is the usual complex conjugation, $\mathcal{M}(\mathbb{R}^{2n})$ is a suitable multiplier space of the space of Schwartz functions $\mathcal{S}(\mathbb{R}^{2n})$ and the star-product is
\begin{equation}
    (f \star_\theta g)(x) 
    = \frac{1}{(\pi\theta)^{2n}} \int \dd^{2n}y\ \dd^{2n}z\ f(x+y) g(x+z) e^{- 2 i y \Theta^{-1} z},\ \ y \Theta^{-1} z = y^\mu \Theta^{-1}_{\mu\nu} z^\nu
    \label{moyal-product}
\end{equation}
for any Schwartz functions $f,g$, which can be extended by duality and continuity to $\mathcal{M}(\mathbb{R}^{2n})$. $\mathbb{R}^{2n}_\theta$ can be equipped with a trace given by the usual Lebesgue integral and one has $(f \star_\theta g)^\dag = g^\dag \star_\theta f^\dag$, $\int \dd^{2n}x\ (f \star_\theta g)(x)
    = \int \dd^{2n}x\ (g \star_\theta f)(x)
    = \int \dd^{2n}x\ f(x)g(x)$.\\
    
It is useful to recall that $[x^\mu, x^\nu]_\theta = i \Theta^{\mu\nu}$ is covariant under the twisted Poincar\'e-Hopf symmetry characterized by the action of a twisted Poincar\'e Hopf algebra $\mathscr{H}$ on $\mathbb{R}^{2n}_\theta$, namely $\mathbb{R}^{2n}_\theta$, $\triangleright: \mathscr{H} \otimes \mathbb{R}^{2n}_\theta \to \mathbb{R}^{2n}_\theta$, this latter being a $\mathscr{H}$-module algebra. The twist, sometimes called the Moyal twist, is
\begin{equation}
       \twiF 
    = e^{- \frac{i}{2} \Theta^{\mu\nu} \partial_\mu \otimes \partial_\nu},\ \twiF \in\mathscr{H}\otimes\mathscr{H}\label{moyal-twist}
   \end{equation}
and the twisted Hopf algebra is characterized by the twisted coproduct and twisted antipode (co-unit unchanged) $ \Delta^\twiF(h)
    = \twiF \Delta(h) \twiF^{-1},\ S^\twiF(h)
    = \chi S(h) \chi^{-1}, \ \chi = \twiF_1 S(\twiF_2)$ ($\twiF=\twiF_1\otimes\twiF_2$) for any $h\in\mathscr{H}$, where $\Delta$ and $S$ are the coproduct and antipode of the usual Hopf Poincar\'e algebra.\\
    
To obtain \eqref{moyal-product}, consider first the 2-d case, i.e. $[x^1, x^2]= iZ$, and simply use the convolution algebra machinery given in subsection \ref{section21} applied to the Heisenberg group $\mathbb{H}_3$, leading to the following convolution product
\begin{equation}
    (f\circ g)(Z,U,V)
    = \int_{\mathbb{R}^3} \dd z \dd u \dd v\ f(z, u, v) g(Z - z + \frac{1}{2}(Uv - Vu), U - u, V - v),
    \label{convol-moyal}
\end{equation}
for any $f,g\in L^1(\mathbb{R}^3)$. Use the map $\#:L^1(\mathbb{R}^3) \to L^1(\mathbb{R}^2)$, $f^\#(u, v)
    = \int_\mathbb{R} \dd z\ f(z, u, v) e^{- i 2 \pi \theta z}$ to define the "twisted convolution product" \cite{vonNeum}
\begin{equation}
    (f\circ g)^\#(u, v)
    = (f^\# \hat{\circ} g^\#)(u, v).
    \label{twist-vonneuman}
\end{equation}
Finally, assume that $f^\#,g^\#$ are momentum space functions and use the Weyl quantization map to obtain $f \star_\theta g
    = \mathcal{F}^{-1}(\mathcal{F}f \hat{\circ} \mathcal{F}g)
    \nonumber$ which gives a 2-d version of \eqref{moyal-product}. The extension to $2n$-dimensions is straightforward. Alternatively, the star-product $\star_\theta$ can be obtained from the Moyal twist \eqref{moyal-twist} since $\mathbb{R}^{2n}_\theta$  is a $\mathscr{H}$-module algebra so that $ h\triangleright\mu_\star(f(x)\otimes g(x))=\mu_\star( \Delta^\twiF(h)\triangleright(f(x)\otimes g(x))$ which leads to
 \begin{equation}
f\star_\theta g=\mu\circ\twiF^{-1}\triangleright(f(x)\otimes g(x))\label{moyal-expans}
\end{equation}
where $\mu_\star(f(x)\otimes g(x))=(f\star_\theta g)(x)$,  $\mu(f(x)\otimes g(x)=f(x)g(x)$, expressing the star-product as an expansion.\\

The (restricted) derivation-based differential calculus underlying (most of) the gauge theories on $\mathbb{R}^{2n}_\theta$ can be easily characterized from the material given in subsection \ref{section22}, with the abelian Lie algebra of derivations $\mathcal{D}=\{\partial_\mu,\ \mu=1,..., 2n \}$. For further use, it is important to notice that
\begin{align}
    \partial_\mu f
    &= [\xi_\mu, f]_\theta, &
    \xi_\mu 
    &= -i \Theta_{\mu\nu}^{-1} x^\nu,
    \label{moyal-innerderiv}
\end{align}
so that any derivation can be formally expressed as an inner derivation. This will be used to define an alternative field variable entering the construction of matrix models in the next subsection.\\
Assuming that $\modM\simeq\mathbb{R}^4_\theta$, $h(m_1,m_2)=m_1^\dag\star m_2$,  setting $\nabla_{\partial_\mu}:=\nabla_\mu$ and 
$\nabla_\mu(\bbone)=-iA_\mu$, so that $A_\mu^\dag
    = A_\mu$ (thus making contact with usual physicists conventions) and choosing $A_\mu$ as the field variable for the gauge theory in the rest of this subsection, one easily arrives \cite{cawa} at
\begin{align}
    \nabla_\mu(f)
    = \partial_\mu f - i A_\mu \star_\theta f,\ i F_{\mu\nu}
    = \partial_\mu A_\nu - \partial_\nu A_\mu - i [A_\mu, A_\nu]_\theta
    \label{moyal-covderiv},
\end{align}
with the group of "unitary gauge transformations"
\begin{equation}
    \mathcal{U}(\mathbb{R}^4_\theta)
    = \{g \in \modM \simeq \mathbb{R}^4_\theta,\ g^\dag \star_\theta g = g \star_\theta g^\dag =\bbone \}
    \label{moyal-gaugegroup},
\end{equation}
and gauge transformations
\begin{equation}
   A^g_\mu 
    = g^\dag \star_\theta A_\mu \star_\theta g + i g^\dag \star_\theta \partial_\mu g,\ \  F^g_{\mu\nu}
    = g^\dag \star_\theta F_{\mu\nu} \star_\theta g.
\end{equation}
Moving now to 4 dimensions, the prototype gauge invariant action is simply given by
   \begin{equation}
    S_{\mathrm{cl}}(A_\mu)
    = \frac{1}{g^2} \int \dd^4x\ (F_{\mu\nu} \star_\theta F^{\mu\nu})(x)\label{moyal-qed}
\end{equation}
and is a mere noncommutative analog of pure QED which has been the subject of many works. Among them, some phenomenological approaches attempted to obtain some 
phenomenological predictions/estimates of what would be the possible imprint of a 
noncommutative QED coupled to fermions (e.g. through $\nabla_\mu(f)
    = \partial_\mu f - i A_\mu \star_\theta f$) resulting in not very stringent lower bounds on the so-called "noncommutativity scale", $\Lambda_{NC}=\frac{1}{\sqrt\theta}$. See e.g. \cite{schupp-a}, \cite{schupp-b} and references therein.\\
The one-loop behavior of \eqref{moyal-qed} has been investigated by using the standard liturgy of the BRST machinery. A gauge-fixed action in a Lorentz-type gauge is
\begin{equation}
    S
    = S_{\mathrm{cl}}(A_\mu) + s \int \dd^4x\ (\overline{C} \partial_\mu A^\mu + \frac{\lambda}{2} \overline{C} b).
    \label{moyal-gaugefixed}
\end{equation}
The BRST operation is $s A_\mu
    = \partial_\mu C - i [A_\mu, C]_\theta, 
    s C
    = \frac{i}{2} [C,C]_\theta, 
    s\overline{C}
    = b, 
    s b
    = 0$ and $s^2=0$.\\
Standard computations show that the 1-point function tadpole is zero as obvious properties of $AAA$ and $\overline{C}CA$ vertices). Unfortunately, the 2-point function, i.e. vacuum polarization $\omega^{\mu\nu}$, exhibits an IR singularity $\sim 1/p^\alpha$ signaling that UV/IR mixing \cite{matusis} does occur in this gauge field theory. In a $d$-dimensional case, the IR limit of $\omega^{\mu\nu}$ is\\
\begin{equation}
    \omega^{\mu\nu}(p)
    = (d-2)\ \Gamma \left( \frac{d}{2} \right) \frac{\tilde{p}^\mu \tilde{p}^\nu}{\pi^{d/2} (\tilde{p}^2)^{d/2}} + \dots,\ \ \tilde{p}^\mu
    = \Theta^{\mu\nu} p_\nu
    \label{moyal-polaris},
\end{equation}
where $\Gamma(z)$ is the Euler Gamma-function which satisfies the transversality condition linked to the Slavnov-Taylor identities: $ p_\mu \omega^{\mu\nu}(p)= 0$, \cite{slavnovjc}. This IR singularity, which does not depend on the gauge choice, cannot be balanced by additional matter contributions. Note that the UV/IR mixing of "$U(N)$" extensions of \eqref{moyal-qed} comes from the pure $U(1)$ part, the $SU(N)$ one being free of mixing.\\
Two types of attempts to get rid of the UV/IR mixing have been investigated but leading to an unsatisfactory net result.\\

The first attempt was based on a clever modification of the above transversality relation, obtained through the addition to \eqref{moyal-qed} of a BF term \cite{slavnov1}:
\begin{equation}
    S
    = S_{\mathrm{cl}}(A_\mu)+\frac{1}{2} \int \dd^4x\ \lambda \star_\theta \Theta^{\mu\nu} F_{\mu\nu},\ \ \Theta_{0j} = 0,\ j = 1, 2, 3
    \label{moyal-bf},
\end{equation}
where $\lambda(x)$ is the $B$-field of the $BF$ term. Then, one can verify that the polarization tensor of the resulting has still a IR singularity of the
form \eqref{moyal-polaris}, still independent of the choice of the gauge function while the propagator of $A_\mu$ becomes now transverse w.r.t. $\tilde{p}^\mu
    = \Theta^{\mu\nu} p_\nu$, i.e. $\tilde{p}^\mu P_{\mu\nu}(p)=0$. This implies that the IR singularity is neutralized whenever the polarization
tensor is connected to a diagram by a propagator of $A_\mu$.\\
However, this nice IR consequence must be strongly temperated. Indeed, the main observation is that the formal commutative limit of \eqref{moyal-bf} is not a gauge theory, but a scalar field theory. Note that possible new UV divergences may appear if the tensor $\Theta^{\mu\nu} $ in \eqref{moyal-bf}is not of full rank. These are the main drawbacks of the works based on the $BF$-term approach \cite{slavnov2}, following the initial work \cite{slavnov1} on $BF$ addition.\\

The second attempt was based on {\it{adaptations}} of the "IR damping methods" used in scalar field theories on (4-d) Moyal space to neutralize the dangerous IR singularities generating UV/IR mixing. Recall that two "IR damping methods" have been successfully used in scalar theories, based either on the addition of the celebrated harmonic term or adding a "$1/p^2$ counterterm" modifying the behavior of the propagator.\\
For instance, recall that in scalar field theory case, the inclusion of a 
harmonic term in the action, says $\sim\Omega\tilde{x}^\mu \phi \star_\theta \tilde{x}_\mu \phi$, gives the following IR (large $x$) behavior for the scalar propagator $P_H(x,0)
    \sim \frac{e^{\frac{- x^2 \theta}{4 \Omega}}}{x^2}$ which thus decays in the IR regime much faster than the usual scalar propagator $ P(x,0)
    \sim \frac{1}{x^2}$. This strong decay acts as an IR cut-off
which renders harmless the effect of the dangerous IR singularities (see \cite{gw-a}).\\
It is not easy to extend the above scheme to gauge theories due to gauge invariance: $\sim x^2 A_\mu \star_\theta A_\mu$ is not gauge invariant. An interesting proposal \cite{danyblasch3} was to introduce the harmonic term through a suitably chosen BRST-exact gauge-fixing term. The resulting gauge-fixed action is thus invariant under a BRST symmetry, hence trading the gauge invariance for a BRST invariance: 
\begin{equation}
  S_H
  = \int \dd^4x\ 
  \frac{1}{4} F_{\mu\nu} \star_\theta F^{\mu\nu}
  + s \left({{\frac{\Omega^2}{8} \tilde{c}_\mu \star_\theta \mathcal{C}^\mu }}+ \overline{C} \star_\theta \partial^\mu A_\mu - \frac{1}{2} \overline{C} \star_\theta b \right)
  \label{moyal-harmonicgauge},
\end{equation}
\begin{equation}
    \mathcal{C}^\mu
    = \{ \{\tilde{x}^\mu, A^\nu\}_\theta, A_\nu\}_\theta
    + [\{\tilde{x}^\mu, \overline{C}\}_\theta, C]_\theta
    + [\overline{C}, \{\tilde{x}^\mu, C\}_\theta]_\theta\label{moyal-grandc},
\end{equation}
with $ s \tilde{c}^\mu= \tilde{x}^\mu$ and other s-transformations as before. Note that the $\tilde{c}_\mu$ vertex does not contribute to loop corrections.
The propagators for $A_\mu$ and ghosts have now the desired IR behaviour recalled just above, namely (in obvious notations):
\begin{align}
    P_{\mu\nu}(p,k)
    &= \delta_{\mu\nu} P_H(p,k), &
    P_{\mathrm{ghost}}(p,k)
    &= P_H(p,k)\label{moyal-gaugepropag}.
\end{align}
However, this proposal suffers from some problems \cite{Blaschke_2010}. One is the loss of transversality for the 2-point function for $A_\mu$, unlike the commutative case, stemming from the breaking of translation invariance due to the choice of $\mathcal{C}^\mu$ \eqref{moyal-gaugepropag}. Indeed, from the Slavnov-Taylor functional identity for \eqref{moyal-harmonicgauge}, one easily obtains
\begin{equation}
   \partial^y_\mu \frac{\delta^2 \Gamma}{\delta A_\nu(x) \delta A_\mu(y)}
    = \int \dd^4z\ \tilde{z}^\mu \frac{\delta^3 \Gamma}{\delta C(y) \delta A_\nu(x) \delta\tilde{c}^\mu(z)}
    \ne 0 .
\end{equation}
Besides, there is a non vanishing tadpole for $A_\mu$ indicating that the vacuum is unstable against radiative corrections. Moreover, the polarization tensor $\omega_{\mu\nu}$ has severe one-loop UV divergences.\\

The adaptation of the $1/p^2$ addition (see \cite{gw-b}) to the gauge theory context has been achieved, resulting in a complicated action with acceptable one-loop properties but the needed modifications appear so far artificial. Besides, nothing about the perturbative behavior is known beyond one-loop. For a review, see \cite{autro-revue}.

\subsection{Gauge matrix models on Moyal spaces}\label{section32}

Recall \eqref{moyal-innerderiv}: $\partial_\mu a
    = [\xi_\mu, a]_\theta$ with $\xi_\mu 
    = -i \Theta_{\mu\nu}^{-1} x^\nu$. This translates into $\dd a
    = [i \xi, a]_\theta
    = i(\xi \star a - a \star \xi)$ where
    $\xi \in {\Omega}^1(\mathbb{R}^{2n}_\theta)$. Then, it is easy to prove \cite{cawa} that there exists a {\it{gauge-invariant connection}} defined for any $a\in\mathbb{R}^{2n}_\theta$ by 
\begin{equation}
   \nabla^{\mathrm{inv}}(a)
    = \dd a - i \xi \star a,\ \ \iff\ \nabla^{\mathrm{inv}}_\mu(a)=\partial_\mu a
    +i\Theta^{-1}_{\mu\nu}x_\nu\star_\theta a 
\end{equation}  
and one verifies for any $g\in\mathcal{U}(\mathbb{R}^{2n}_\theta)$ and any $a \in \algA$ that $ (\nabla^{\mathrm{inv}})^g(a)
    = \nabla^{\mathrm{inv}}(a)$. It is then natural to consider the difference of two connections $\nabla - \nabla^{\mathrm{inv}}$, which thus defines a covariant tensor 1-form $\mathcal{A} \in {\Omega}^1(\mathbb{R}^{2n}_\theta)$ with
\begin{equation}
 \mathcal{A}_\mu
    = - i (A_\mu + {{\Theta^{-1}_{\mu\nu} x^\nu}})  
    \label{moyal-cov-coord}
\end{equation}
and, as such, transforms covariantly under $\mathcal{U}(\mathbb{R}^{2n}_\theta)$, namely 
\begin{equation}
    \mathcal{A}_\mu^g
    = g^\dag \star \mathcal{A}_\mu \star g
    \label{moyal-gaugearond}.
    \end{equation}
This is the "covariant coordinate" of the String physics literature. The tensor form coordinate will be the field variable chosen in this subsection instead of the (noncommutative) gauge potential used in the subsection \ref{section31}. Expressed in this variable, the curvature takes the form\\
\begin{equation}
    F_{\mu\nu}
    = [\mathcal{A}_\mu, \mathcal{A}_\nu]_\theta - i \Theta^{-1}_{\mu\nu}\label{moyal_Fcovcoord},
\end{equation}
where the curvature for the invariant connection is nothing but the last term in \eqref{moyal_Fcovcoord}, namely $ F^{\mathrm{inv}}_{\mu\nu}=- i \Theta^{-1}$.\\
In \cite{wall-wulk1} has been singled out a 4-d gauge invariant action of the following form
\begin{equation}
    S(\mathcal{A})
    = \int \dd^4 x\ \left(
    - \frac{1}{4} [\mathcal{A}_\mu, \mathcal{A}_\nu]_\theta^2
    + \frac{\Omega^2}{4} \{{\cal{A}}_\mu, {\cal{A}}_\nu\}^2_\theta
    + \kappa {\cal{A}}_\mu \star_\theta {\cal{A}}^\mu \right)
    \label{moyal_ikkt}
\end{equation}
where $\Omega$ and $\kappa$ are constants, which involves obviously a term $\sim F_{\mu\nu} \star_\theta F^{\mu\nu}$ in view of \eqref{moyal_Fcovcoord} which is necessary in order to obtain a suitable commutative limit for \eqref{moyal_ikkt}. This gauge invariant actions can be viewed as describing a family of matrix models, reminiscent of the type IIB matrix models \cite{matrix1-a}, \cite{matrix1-b}. \\
Note that the initial (naive) expectation was that the term $\sim\{{\cal{A}}_\mu, {\cal{A}}_\nu\}^2_\theta$ could play the role of some harmonic term as it involves terms $\sim x^2A_\mu^2$ while preserving the gauge invariance. Unfortunately, the gauge theory \eqref{moyal_ikkt} has a complicated vacuum structure \cite{wall-wulk2} which preclude practical computations to be carried out beyond the classical order. In fact, $A_\mu=0$ is not a vacuum while the trivial $\mathcal{A}_\mu=0$ vacuum obviously yields a non dynamical matrix model.\\

Solving the equation of motion for \eqref{moyal_ikkt}
\begin{equation}
   0
   = - 2 (1 - \Omega^2) \mathcal{A}^\nu \star_\theta \mathcal{A}_\mu \star_\theta \mathcal{A}_\nu
   + (1 + \Omega^2) \mathcal{A}_\mu \star_\theta \mathcal{A}^\nu \star_\theta \mathcal{A}_\nu \\
   + (1 + \Omega^2) \mathcal{A}^\nu \star_\theta \mathcal{A}_\nu \star_\theta \mathcal{A}_\mu
   + 2 \kappa \mathcal{A}_\mu
   \label{moyal-motion},
\end{equation}
has been carried out in \cite{wall-wulk2} using the matrix base of $\mathbb{R}^4_\theta$ and $\mathbb{R}^2_\theta$. Recall that in 2-d case{\footnote{The extension to the 4-d case is straightforward.}}, the set of eigenfunctions of the 1-d harmonic oscillator, says $\{f_{mn}(x) \}$$_{m,n\in\mathbb{N}}$, is an orthogonal 
basis of $\mathcal{S}(\mathbb{R}^2)$ so that $a(x) = \sum_{m,n} a_{mn} f_{mn}(x)$ for any $a\in\mathcal{S}(\mathbb{R}^2)$, with
\begin{equation}
    f_{mn} \star_\theta f_{kl}
    = \delta_{nk} f_{ml}, \qquad
    f_{mn}^\dag
    = f_{nm},\langle f_{mn}, f_{kl} \rangle_{L^2} 
    = \int \dd^2x\ (f_{mn}^\dag \star_\theta f_{kl})(x)
    = 2 \pi \theta \delta_{mk} \delta_{nl}\label{eq:moyal_matbas}.
\end{equation}
This base is particularly useful to formulate matrix models as in particular the star-product $\star_\theta$ reduces to a matrix product, namely $(a \star_\theta b)(x)
    = \sum_{m,n\in\mathbb{N}} {{\left( \sum_{k\in\mathbb{N}} a_{mk} b_{kn} \right)}}  f_{mn}(x)$.
This can be understood by noticing that the algebra $(\mathcal{S}(\mathbb{R}^2),\star_\theta)$ is isomorphic to the algebra $\mathbb{M}_\theta \subset \ell^2(\mathbb{N}^2)$, the subalgebra of $\ell^2(\mathbb{N}^2)$ involving rapid decay matrices.
The isomorphism and inverse are $\phi_{mn} 
    \mapsto \sum_{m,n} \phi_{mn} f_{mn} \in \mathcal{S}(\mathbb{R}^2),\  \phi \in \mathcal{S}(\mathbb{R}^2)
    \mapsto \frac{1}{2 \pi \theta} \langle \phi, f_{mn} \rangle_{L^2}$. In fact, any element of $\mathbb{M}_\theta$, $\phi=\sum_{m,n} \phi_{mn}e_m\otimes e_n$ where $(e_n)_{n\in\mathbb{N}}$ is a basis of $\ell^2(\mathbb{N})$ with dual basis $(e'_n)_{n\in\mathbb{N}}$ ($e'_n(e_p) = \delta_{np}$), can be identified with an operator of $\ell^2(\mathbb{N})$ by making use of the faithful representation $\eta: \ell^2(\mathbb{N}) \otimes \ell^2(\mathbb{N}) \to {\cal{B}}(\ell^2(\mathbb{N}))$, $\eta(e_m \otimes e_n) 
    = e_m \otimes e'_n$, for any $m,n\in\mathbb{N}$. One obtains the operator $ \widehat{\phi}
    = \sum_{m,n} \phi_{mn} e_m \otimes e'_n$, or in physicist notation $\widehat{\phi}
    =\sum_{m,n} \phi_{mn} |m\rangle \langle n|$. In this framework, one thus trades any function $\phi(x)= \sum_{m,n} \phi_{mn} f_{mn}(x)$ for an 
    operator $\widehat{\phi}
    =\sum_{m,n} \phi_{mn} |m\rangle \langle n|$ and the $f_{mn}(x)$ are the symbols of the operators $|m\rangle \langle n|$.\\

Solutions of the equation of motion \eqref{moyal-motion}, invariant under $SO(d)\cap Sp(d)$ have been obtained by tedious computation \cite{wall-wulk2} using the above matrix base. It turns out that their complicated expressions in the 4-d case prevent their use in practical field theory computation beyond the 
classical order. \\
One particular configuration in the 2-d case has however been exploited in \cite{MVW13} in an attempt to explore one-loop properties of a 2-d gauge matrix model. Unfortunately, it appears that the 1-loop tadpole function for the field is not zero, signaling a vacuum instability against quantum corrections.

\subsection{Gauge models on $\mathbb{R}^3_\lambda$}\label{section33}

The deformations of $\mathbb{R}^3$, called generically $\mathbb{R}^3_\lambda$, are quantum spaces with $\mathfrak{su}(2)$ coordinates Lie algebra written as $[x^j, x^k]
    = i \lambda \tensor{\epsilon}{^{jk}_l} x^l$, where $\lambda$, the deformation parameter, has dimension of a length. \\
    In the spirit of subsection \ref{section21}, a convenient star-product can be obtained by considering the convolution
algebra of $\SU$. Since $\SU$ is compact, the Peter--Weyl theorem
applies (for an introduction to harmonic analysis on compact groups, see
\cite{fritz}). It provides a Hilbert-space decomposition of
$L^2(\SU)$ into irreducible unitary representations of $\SU$. Denoting by $\pi_j$ the irreducible unitary representation
of spin $j$, with representation space $V_j$ and dimension $d_j=2j+1$,
one has
\begin{equation}
L^2(\SU)
\simeq
\widehat{\bigoplus}_{j\in\frac{\NN}{2}}
\End(V_j)
\simeq
\widehat{\bigoplus}_{j\in\frac{\NN}{2}}
\mathbb{M}_{2j+1}(\CC),
\label{eq:r3l_PW_isomorp1}
\end{equation}
where the hat denotes the Hilbert direct sum with the
Plancherel norm $\|f\|_{L^2(\SU)}^2
=
\sum_{j\in\frac{\NN}{2}}
(2j+1)\,
\|\widehat f(j)\|_{\HS}^2$
in which $\|\cdot\|_{\HS}$ is the Hilbert--Schmidt norm and
$\widehat f(j)$ is the Fourier transform of $f$ in the representation
$\pi_j$. Upon normalizing the Haar measure on $\SU$ to $1$ and introducing the Fourier transform $\widehat f(j)=\int_{\SU}\dd x\, f(x)\pi_j(x)$, one obtains that the Fourier transform maps the convolution product $(f*g)(x)=\int_{\SU}\dd y\, f(y)g(y^{-1}x)$ into the usual matrix product $\widehat{f*g}(j)=\widehat f(j)\widehat g(j)$, while the involution $f^*(x)=\overline{f(x^{-1})}$
is mapped into the usual Hermitian conjugation of the Fourier matrices. Let $\{e_m^j\}_{-j\leq m\leq j}$ be an orthonormal basis of $V_j$ and $\pi^j_{mn}(x)$ the matrix coefficients of the representation $\pi_j(x)$ in this basis. They obey the orthogonality relations $\int_{\SU}\dd x\,
\overline{\pi^{j_1}_{mn}(x)}
\pi^{j_2}_{pq}(x)
=
\frac{1}{2j_1+1}
\delta_{j_1j_2}\delta_{mp}\delta_{nq}$. It follows that any $f\in L^2(\SU)$ admits the following Peter--Weyl expansion given by $f(x)=\sum_{j\in\frac{\NN}{2}}
(2j+1)\operatorname{Tr}\!\left(\widehat f(j)\pi_j(x)^\dagger\right)$. Hence, the Fourier transform is an isomorphism of Hilbert spaces
between $L^2(\SU)$ and the weighted Hilbert direct sum appearing in the relation \eqref{eq:r3l_PW_isomorp1} displayed above. For our present purpose, it will be convenient to consider the algebraic direct sum
\begin{equation}
\mathcal A
=\bigoplus_{j\in\frac{\NN}{2}}^{\mathrm{alg}}
\mathbb{M}_{2j+1}(\CC),
\label{eq:r3l_alg_sum}
\end{equation}
whose elements are sequences of matrices with only finitely many
non-vanishing components. This algebraic direct sum is a dense
$*$-subalgebra of the Hilbert-space completion in
Eq.~\eqref{eq:r3l_PW_isomorp1}. The related product and involution are defined blockwise by usual matrix multiplication and Hermitian conjugation. Through the Fourier transform, $\mathcal A$ can
equivalently be identified with the subalgebra of $L^2(\SU)$ whose elements are functions with finite Peter--Weyl expansions.\\

Making contact with the notations of the physics literature can be achieved by introducing $v^j_{mn}:=
|jm\rangle\langle jn|
\equiv
e_m^j\otimes e_n^{j*},
\ 
j\in\frac{\NN}{2},\ -j\leq m,n\leq j$,  where $\{e_n^{j*}\}$ denotes the dual basis of $V_j^*$, defined by
$e_n^{j*}(e_p^j)=\delta_{np}$. These form the canonical
basis for $\mathbb{M}_{2j+1}(\CC)$ and satisfy $(v^j_{mn})^\dagger=v^j_{nm}$ and $v^{j_1}_{mn}v^{j_2}_{pq}
=
\delta_{j_1j_2}\delta_{np}\,
v^{j_1}_{mq}$ with $-j_1\leq m,n\leq j_1$ and $-j_2\leq p,q\leq j_2$. Thus, any element $a\in\mathcal A$ can be written as
\begin{equation}
a
=
\sum_{j\in\frac{\NN}{2}}
\sum_{-j\leq m,n\leq j}
a^j_{mn}v^j_{mn},
\qquad
a^j_{mn}\in\CC,
\label{eq:r3-decompos}.
\end{equation}
We are now in position to define the noncommutative algebra
$\mathbb{R}^3_\lambda$ as the following $*$-algebra:
\begin{equation}
\mathbb{R}^3_\lambda
=
\left(
\bigoplus_{j\in\frac{\NN}{2}}^{\mathrm{alg}}
\mathbb{M}_{2j+1}(\CC),
\cdot,\dagger
\right).
\label{r3-alg}
\end{equation}
The associative product is the usual matrix product applied
independently in each representation block, while the involution is
the usual Hermitian conjugation. The Hilbert-space completion inherited from the Peter--Weyl decomposition is understood with the Plancherel weights $2j+1$ appearing in the Plancherel norm introduced above. I will use this description in the rest of this subsection.\\
In the language of physicists, $ \mathbb{R}^3_\lambda $ is described as an infinite (direct) sum of fuzzy spheres. Orthogonality holds w.r.t the Hilbert product $\langle f,g \rangle := \tr(f^\dag g)$ with the trace given by 
\begin{equation}
    \tr(f \star_\lambda g) 
    := 8 \pi \lambda^3 \sum_{j\in\frac{\mathbb{N}}{2}} (2j+1) \tr_j(F^j G^j)
    \label{eq:r3l_trace},
\end{equation}
for any $f,g\in\mathbb{R}^3_\lambda$, where $F^j, G^j \in \mathbb{M}_{2j+1}(\mathbb{C})$ are the matrix arising in the blockwise expansion of $f, g$ respectively in the canonical basis and $\tr_j$ is the usual trace on $\mathbb{M}_{2j+1}(\mathbb{C})$. The overall factor $8 \pi \lambda^3$ in \eqref{eq:r3l_trace} has been set for convenience. \\

The derivation-based differential calculus can be straightforwardly characterized from the material of subsection \ref{section22} using the Lie algebra of real inner derivations of $\mathbb{R}^3_\lambda$ \cite{gervitwal}:
\begin{align}
    \mathcal{D} := \{D_\alpha:= i [\theta_\alpha, \cdot]_\lambda\}, 
    \theta_\alpha := \frac{x_\alpha}{\lambda^2},\ [D_\alpha, D_\beta] = -\frac{1}{\lambda} \tensor{\epsilon}{_{\alpha\beta}^\gamma} D_\gamma
    \label{r3-derivlie}.
\end{align}
One further assumes as before that the right module used to define the connection is one copy of $\mathbb{R}^3_\lambda$ with hermitian structure $h(m_1, m_2) = m_1^\dag \star_\lambda m_2$. The hermitian connection and curvature are easily find to be
\begin{align}
    \nabla_{D_\mu}(f) 
    &:= \nabla_\mu(f) 
    = D_\mu (f) + A_\mu \star_\lambda f, & 
    (A_\mu :=\nabla_\mu(\bbone))
    \label{eq:r3l_conn}
\end{align}
and 
\begin{equation}
    A_\mu^\dag = - A_\mu, 
\end{equation}
for any $f\in\mathbb{R}^3_\lambda$ and \cite{gervitwal}
\begin{align}
\begin{aligned}
    F_{\mu\nu} 
    = D_\mu (A_\nu) - D_\nu (A_\mu) + [A_\mu,A_\nu]_\lambda + \frac{1}{\lambda} \tensor{\epsilon}{_{\mu\nu}^\gamma} A_\gamma.
    \mu, \nu = 1,2,3.
\end{aligned}
    \label{eq:r3l_curv}
\end{align}
The gauge group is the group of the unitary elements of the module $\mathcal{U}(\mathbb{R}^3_\lambda)$, i.e. such that $g := \phi(\bbone),
    \phi \in \Aut(\modM, h)$, $g^\dag \star_\lambda g = g \star_\lambda g^\dag = \bbone $ and one has
\begin{align}
    A_\mu^g = g^\dag \star_\lambda A_\mu \star_\lambda g + g^\dag \star_\lambda D_\mu (g), &&
    F^g_{\mu\nu} = g^\dag \star_\lambda F_{\mu\nu} \star_\lambda g.
\end{align}
In view of the Lie algebra of derivation $\mathcal{D}$ \eqref{r3-derivlie}, there exists again a gauge invariant connection. One verifies that it is given by 
\begin{equation}
    \nabla^{\mathrm{inv}}(f)
    := \dd f + \Theta \star_\lambda f
    = f \star_\lambda \Theta,\ \Theta(D_\mu):=\Theta_\mu=-i\theta_\mu
    \label{eq:r3l_inv_form_conn},
\end{equation}
for any $ f\in\mathbb{R}^3_\lambda$ where $\theta_\mu$ given in \eqref{r3-derivlie}
and $\Theta_\mu^g
    = \Theta_\mu$ with $ F^{\mathrm{inv}}
    := \dd \Theta + \Theta \star_\lambda \Theta
    = 0$ so that the invariant connection is a flat connection. The covariant coordinate is now \cite{gervitwal}
\begin{equation}
    \mathcal{A}_\mu 
    = \nabla_\mu - \nabla^{\mathrm{inv}}_\mu 
    = A_\mu + i \theta_\mu\label{r3-covcoord},
\end{equation}   
and the curvature can be re-expressed in this field variable as 
\begin{equation} 
    F_{\mu\nu} 
    = [\mathcal{A}_\mu, \mathcal{A}_\nu]_\lambda + \frac{1}{\lambda} \tensor{\epsilon}{_{\mu\nu}^\gamma} \mathcal{A}_\gamma.
    \label{r3-vraicurv}
\end{equation}
As for the Moyal case, the choice of the type of field variable, $A_\mu$ or $\mathcal{A}_\mu$, leads to different gauge theories, respectively an analog of a Yang-Mills theory for $A_\mu$ or a gauge invariant matrix model for $\mathcal{A}_\mu$.\\

Consider first $A_\mu$ as field variable. Then, looking for a gauge invariant action functional at most quartic in $A_\mu$, involving no linear terms in $A_\mu$ and with a positive kinetic operator, and using $\mathcal{A}_\mu
    = \sum_{j\in\frac{\mathbb{N}}{2}} \sum_{-j \leqslant m, n\leqslant j} (\mathcal{A}_\mu^j)_{mn} v^j_{mn}$ and the properties of the matrix base introduced at the beginning of the subsection, one arrives at an action of the form $S_{\mathrm{cl}}[A] 
    = \sum_{j \in \frac{\mathbb{N}}{2}} S^{(j)} [A]$ with 
\begin{equation}
    S_{\mathrm{cl}}(A_\mu)
    = \frac{8 \pi \lambda^3}{g^2} \sum_{j \in \frac{\mathbb{N}}{2}}(j+1)
    \Big( (F^j_{\mu\nu})^\dag F^j_{\mu\nu}
    +\gamma \big( \epsilon^{\mu \nu \rho} \mathcal{A}^j_\mu \mathcal{A}^j_\nu \mathcal{A}^j_\rho
    + \frac{3}{2\lambda} \mathcal{A}^j_\mu (\mathcal{A}^j)^\mu \big) \Big),\ \mathcal{A}_\mu=\mathcal{A}_\mu(A)\label{r3-actionym},
\end{equation}
which is expressed as an infinite sum of actions $S^{(j)}$, each describing describing a Yang-Mills-Chern-Simons actions on a fuzzy spheres $\mathbb{S}^j \simeq \mathbb{M}_{2j+1}(\mathbb{C})$. Notice that $S^{(j)}$ is similar to the action proposed in \cite{ARS-a} (see also \cite{ARS-b}, \cite{ARS-c}) and describing the dynamics of open strings in a curved space with the metric of a 3-sphere in the presence of a  non-vanishing Neveu-Schwarz $B$ field and with $D$ brane. Using the matrix base, loop computations can be done rather easily. Unfortunately, the 1-point tadpole function for $A_\mu$ is found to be non vanishing indicating that the classical vacuum is not stable against quantum fluctuations. Besides, UV/IR mixing shows up.\\

The situation is much more interesting when $\mathcal{A}_\mu$ is chosen as field variable.\\
Observe first that $\tr[(P(\mathcal{A}) \Theta_\mu \Theta^\mu)^g]
    = \tr(P(\mathcal{A}) \Theta_\mu \Theta^\mu)$ stemming from the properties: $(\Theta^\mu)^g=\Theta^\mu$ and $\Theta_\mu\Theta^\mu\in\mathcal{Z}(\mathbb{R}^3_\lambda)$, where $P(\mathcal{A})$ is any polynomial depending on $\mathcal{A}$ and $\Theta$ is defined in \eqref{eq:r3l_inv_form_conn}. This implies that gauge invariant harmonic terms $\sim \tr(x^2 \Phi_\mu \Phi^\mu)$
are now allowed in the gauge action.\\
Changing the notation as $\mathcal{A}_\mu \to \Phi_\mu$ and keeping the same assumptions on the classical action $S(\mathcal{A})$ as those for $S(A)$, standard algebraic computations yield \cite{wal-16}
\begin{equation}
    S_{\mathrm{cl}}[\Phi]
    = \frac{1}{g^2} \tr \Big( 
    [\Phi_\mu, \Phi_\nu]^2 + \Omega \{\Phi_\mu, \Phi_\nu\}^2 + (M + \mu x^2) \Phi_\mu\Phi^\mu \Big),
    \label{decadix}
\end{equation}
which supports $\Phi_\mu=0$ as classical vacuum and where $\Omega$, $\mu$ and $M$ are constants.\\
A standard BRST gauge-fixing, using the gauge $\Phi_3=\theta_3$, leads to
\begin{equation}
    S^f_\Omega 
    = \frac{2}{g^2} \tr\big( \Phi \mathcal{Q} \Phi^\dag + \Phi^\dag \mathcal{Q}\Phi \big)
    + \frac{16}{g^2} \tr \big( (\Omega+1) \Phi \Phi^\dag \Phi \Phi^\dag + (3 \Omega - 1) \Phi \Phi \Phi^\dag \Phi^\dag \big),
    \label{decadix-gfaction}
\end{equation}
\begin{equation}
    \mathcal{Q}
    = M \bbone + \mu L(x^2) + 8 \Omega L(\theta_3^2) + 4 i (\Omega - 1) L(\theta_3) D_3
    \label{decadix-q},
\end{equation}
where $\Phi := \frac{1}{2}(\Phi_1 + i \Phi_2)$.\\
The expression for the gauge fixed action \eqref{decadix-gfaction}, \eqref{decadix-q} is close to the one describing the family of complex LSZ models \cite{LSZ-a}, \cite{LSZ-b} the main difference being the kinetic operators in both actions. Recall that LSZ models are noncommutative complex scalar field theories on $\mathbb{R}^{2n}_\theta$, some being exactly solvable, stemming from a duality between space coordinates and momenta which appears
when a harmonic term is present \cite{LSZ-a}. \cite{LSZ-b}.\\

When $\Omega=\frac{1}{3}$, the action described by \eqref{decadix-gfaction}, \eqref{decadix-q} is
formally similar to the action for an exactly solvable LSZ model (up to differences
for the respective kinetic operators). This gauge theory model is exactly
solvable as shown in \cite{bibi}, where exact solvability must be understood in a blockwise sense as explained now. \\
Indeed, the corresponding partition function is expressible as a product of factor, stemming from the Peter-Weil decomposition of $\mathbb{R}^3_\lambda$
\begin{equation}
Z(Q)=\prod_{j\in\frac{\mathbb{N}}{2}}Z_j(Q),\ \   Z_j(Q) = (N^j(g^2) \ (2j+1)!)\frac{\det_{-j\le m,n\le j} \left(f(\omega^j_m+\omega^j_n)\right)}{\Delta^2(Q^j)} \  \  \ , \label{decadix-Z}
\end{equation}
\begin{equation}
f(x) = \sqrt{\frac{\pi g^2}{128(j+1)}} \ \ \text{erfc}(x\sqrt{\frac{(j+1)}{64g^2}}) \ \ e^{x^2\frac{(j+1)}{64g^2}}\  \label{decadix-erfc}
\end{equation}
with 
\begin{equation}
   N^j(g^2)=(\prod_{k=1}^{2j}k!)^2(\frac{2w(j)}{g^2})^{-2j(2j+1)},\ \ w(j)=8\pi\lambda^3(2j+1),
\end{equation}
and $\Delta(Q^j)$ is the Vandermonde determinant of the (symmetric real) matrix $Q^j\in\mathbb{M}_{2j+1}(\mathbb{C})$ related to the kinetic operator whose expression obtained from the use of the matrix base is
\begin{equation}
 Q^j_{mn;kl}=\delta_{ml} \delta_{nk} (M+\mu\lambda^2j(j+1))  +\frac{2}{3\lambda^2}(k+l)^2+\frac{2}{\lambda^2}(k-l),
\end{equation}
$\text{erfc}$ is the complementary error function and $\omega^j_m$ are the eigenvalues of $Q^j$. Note that the functional integrals in $Z_j(Q)$ can be carried out so that any corresponding truncated gauge model. Each $Z_j(Q)$ is nothing but a $\tau$-function for a 2-d Toda hierarchy \cite{bibi} and thus can be interpreted as the partition function for the reduction of the gauge-fixed theory on $\mathbb{M}_{2j+1}(\mathbb{C})$, i.e. on a fuzzy
sphere of radius $j$.\\
Therefore, each block (partition function) $Z_j$ for the reduction of the full gauge model on each fuzzy sphere modeled by $\mathbb{M}_{2j+1}(\mathbb{C})$ is exactly solvable. However exploring the exact solvability of the full gauge model would at least amount to study the convergence of the infinite product $Z(Q)=\prod_{j\in\frac{\mathbb{N}}{2}}Z_j(Q)$ (see 1st of \eqref{decadix-Z}) which so far is not known.\\

For $\Omega>0$, the gauge model is perturbatively finite to all orders \cite{wal-16}. The lengthy proof results from a combination of a sufficient rapid decay of the propagator at large $j$, the role of UV and IR cut-off played by $j$ and the existence of an upper bound for the propagator. One thus has:
\begin{th-imp}
[\cite{wal-16}] The amplitudes of the (ribbon) diagrams for any of the gauge theories described by $S^f_{\Omega}(\Phi)$ with $M>0$, $\mu>0$, $\Omega>0$, are finite to all orders in perturbation.
\end{th-imp}
Note that the presence of a harmonic term ($\mu\ne0$) is essential for the above property to hold. \\
This interesting result in itself must however be tempered. The commutative limit of $S^f_{\Omega}(\Phi)$ leads to a non usual (3-d) model. Note that deformations of $\mathbb{R}^3$ has appeared in Group Field Theory developments \cite{Oriti_2009-a}, \cite{Oriti_2009-b}. Group field theory appeared in the context of quantum gravity and aim at modeling quantum gravity from a combinatoric of non-local quantum field theories on group manifolds. While gauge theories on $\mathbb{R}^3_\lambda$ share some features with the noncommutative/matrix model representations of group field theory models \cite{MDFG}, it is so far unknown if (some of) the above  gauge theories may be actually related to particular group field theory models.
\section{Gauge theories on deformations of Minkowski space-time}\label{section4}
Quantum deformations of the Minkowski space-time have received a huge attention. One consensus prevailing is that they are of possible relevance in a description of an effective regime of Quantum Gravity \cite{zerevue}, \cite{whitepaper}. Among these deformations, the $\kappa$-Minkowski space-time is probably the most popular
\cite{luk2}, as providing a realization of the Double Special Relativity \cite{kg-2005} or for its relationship to Relative Locality \cite{rel-loc1}. Other Lie-algebraic deformations of the Minkowski space-time are also known for a long time but have not been intensively exploited so far. These quantum Minkowski space-times are linked to various quantum deformations of the Poincar\'e (Hopf) algebra, these acting as "quantum symmetries" on these space-times. These deformations are in fact linked to the Poisson structures of the Poincar\'e group given by classical $r$-matrices, classified a long time ago \cite{zak}. This gave rise \cite{lukier1} to various Lie-algebraic deformations of the Minkowski space-time with coordinates Lie algebra of the general form:
\begin{equation}[x^\mu,x^\nu]=i\zeta^\mu(\eta^{\mu\beta} x^\alpha-\eta^{\nu\beta} x^\alpha)-i\zeta^\nu(\eta^{\mu\beta} x^\alpha-\eta^{\nu\beta} x^\alpha),
\end{equation}
where $\zeta^\mu$ is a vector with dimension of a length and $\alpha$ and $\beta$ fixed. This analysis has been somehow extended in \cite{mercati-a}, \cite{mercati-b} under reasonable assumptions leading to the characterization of 17 classes of centrally-extended Lie algebras of coordinates defining new quantum Minkowski space-times.\\
A systematic construction of star products and involutions for these models restricted to the case of non centrally-extended Lie algebras of coordinates, thus defining the corresponding $*$-algebras for these quantum Minkowski space-times as been carried out recently in \cite{allstar}. This led to the characterization of 11 new quantum Minkowski space-times, having "noncommutativity of Lie-algebra type", which will be the focus of this subsection. \\

The construction of the star-products and involutions is a mere application of the material recalled in the subsection \ref{section21}. The quantum Minkowski space-times to be characterized have Lie groups $\mathcal{G}$ related to the Lie algebras of coordinates $\mathfrak{g}$ of the form of semi-direct products as $\mathcal{G}
= H\ltimes_{{\phi}} \mathbb{R}^{n}$ where the abelian subgroup $H\subset GL(n,\mathbb{R})$ acts on $\mathbb{R}^n$ as $\phi_a(x)=ax $ for $a\in H$, $x\in\mathbb{R}^n$, $n\geq1$. This is the configuration presented in subection \ref{section21}. The result is that any of these quantum Minkowski space-times can be modeled by an associative $*$-algebra $\mathcal{M}=(\mathbb{C}(\mathcal{G}),\star,\dag)$ with the following star-product and involution:
\begin{equation}
    \begin{split}
        \left(f\star g\right)(x) &= \frac{1}{(2\pi)} \int dp^M dy^M e^{-i p^M y^M} f(x + y^M)g(A(p^M)x),\\
    f^\dag(x)&= \frac{1}{(2\pi)}\int dp^Mdy^M\ e^{-ip^My^M}f(A(p^M)x + y^M)
    |\det(A(p^M))|^2\label{gene-invol-bis},
    \end{split}
\end{equation}
for any $f,g\in\mathcal{M}$ and $a\in H$ where $x^M$ labels the special coordinate which "generates the noncommutativity", namely the coordinate algebra for \eqref{gene-invol-bis}  verifies
\begin{equation}
    \left[x^M,x^\mu\right] = -i \left[\partial_{p^M}A(p^M)\rvert_{p^M=0}\right]^\mu\;_\sigma x^\sigma\;,
\label{general-liecommut}
\end{equation}
with all other commutators being equal to $0$ (and obviously satisfies the Jacobi identity so that \eqref{general-liecommut} defines a Lie algebra). In \eqref{gene-invol-bis}, \eqref{general-liecommut}, the matrix 
$A(p^M)$ is given by
\begin{equation}
    A(p^M) = \mathbb{I}^M \oplus a^T(p^M) \oplus^{4 - n-1} \mathbb{I},
\end{equation}
where the matrix $a$ is defined by the parametrization (faithful representation) of $\mathcal{G}$
\begin{equation}
s(p^M,\vec{p})=\begin{pmatrix} a(p^M) & \vec{p} \\ 0 & 1   \end{pmatrix}\label{decadix112},
\end{equation}
where it is assumed that functions of the group algebra for $\mathcal{G}$ are functions on the momentum space (cf. subsection \ref{section21}). The matrix $a(p^M)$ satisfies $a(p^M)a(q^M)=a(p^M+q^M)$, $a^{-1}(p^M)=a(-p^M)$ and $a(0)=\bbone$, for any $a\in H\subset GL(n,\mathbb{R})$.\\
From standard results on analysis on locally compact groups, $d\nu_\mathcal{G}(s)=\Delta_\mathcal{G}(s^{-1})d\mu_\mathcal{G}(s)$ linking right and left Haar measure, respectively $d\mu_\mathcal{G}$ and $d\nu_\mathcal{G}$, where the modular function $\Delta_\mathcal{G}$ verifies $\Delta_\mathcal{G}(s)
    = |\det(a)|^{-1}$ together with $d\mu_\mathcal{G}((a,x))
    = d^nx\ dz\ |\det(a(z))|^{-1}$ which holds for semi-direct product groups as $\mathcal{G}$, one concludes that the right-Haar measure coincides with the Lebesgue measure. Furthermore, one verifies that\\
\begin{equation}
    \int d^4x( f\star g )= \int d^4x (\left[\det A(P^M)\vartriangleright g\right]\star f),\ \ P_M = -i \frac{\partial}{\partial x^M}.\label{trace-rho}
\end{equation}
It follows that the Lebesgue integral defines a trace for the above star-product when $\det A(P^M)=\det a(P^M)=1$ corresponding to unimodular $\mathcal{G}$, as a cyclic positive map. Indeed, one has
\begin{equation}
    \int dx^4\ (f^\dag\star g)(x)=\int dx^4\ \overline{f}(x)g(x)\label{positivemap},
\end{equation}
implying positivity of the map defined by $\int d^4x$. Whenever $\det A(P^M)\ne1$, the Lebesgue integral actually defines a KMS weight, as it is already the case for $\kappa$-Minkowski case \cite{PW2018}. The corresponding consequences will be discussed in a while.\\

The resulting quantum Minkowski-space-times are collected in the Table 1, giving the expression for the matrix $A$ in each case and displaying separately the unimodular and non-unimodular cases. When $\mathcal{G}$ is unimodular, the cases (13, 14, 15) where $H\simeq SO(2)\ltimes\mathbb{R}^2$ correspond to the so-called $\rho$-Minkowski space-time, discussed in the sequel, while the case 16 is the "hyperbolic version" of $\rho$-Minkowski, with $H\simeq SO(1,1)\ltimes\mathbb{R}^2$. For the cases 10 and 12, $\mathbb{H}$ is the Heisenberg group. For the properties of the groups $\mathcal{G}$ appearing for the non-unimodular case, see \cite{lieclassif}.\\

In the following, I will focus only on the (unimodular) cases 13, 14, 15 of table 1 which all correspond to the $\rho$-Minkowski space-time. This quantum space-time has been considered recently in \cite{marija2-a}, \cite{marija2-b}. The construction of a gauge theory on this quantum space-time is based on a twisted derivation-based differential calculus and uses the notion of twisted connection on a right-module which is somehow similar to the one used in the subsection \ref{section31}. The corresponding framework can be easily adapted from subsection\ref{section22} so that I will quote only the useful material. It is also instructive to compare the obtained results to those obtained by applying a similar framework to the case of the popular $\kappa$-Minkowski space-time for which the group $\mathcal{G}=\mathbb{R}\ltimes \mathbb{R}^{d}$ for a "$(d+1)$-dimensional space-time. In this latter case, the Lebesgue integral is no longer a trace, i.e. one has $\det A(P^M)=e^{-dP_0/\kappa}$, $P_0=-i\partial_0$, which thus defines a "twisted trace" in the physicist language which should be more properly called a KMS weight \cite{kuster-a}, \cite{kuster-b}.\\

\subsection{Yang-Mills type theory on $\rho$-Minkowski space-time}\label{section41}

The $\rho$-Minkowski space-time can be viewed as generated by the following Lie algebra of coordinates
\begin{align}
    [x_0, x_1] = i \rho x_2, &&
    [x_0, x_2] = - i \rho x_1, &&
    [x_1, x_2] = 0,
    \label{coord-alg-intro}
\end{align}
where $\rho$ has the dimension of a length which is supplemented by another central generator $x_3$. Note that one could interchange $x_0$ and $x_3$ which would correspond to a physically different situation where the time $x_0$ would stay "commutative". In the following, I will not consider this possibility. The star-product and involution can be obtained from the table 1. One gets (in obvious notations)
\begin{align}
    (f\star_\rho g)(x_0,\vec{x}) 
    &= \int \frac{dp_0}{2\pi}\ dy_0\ e^{-i p_0 y_0} f(x_0 + y_0, \vec{x}) g(x_0, R(-\rho p_0)\vec{x}),
    \label{star-rho} \\
    f^\dag(x_0,\vec{x})
    &= \int \frac{dp_0}{2\pi}\ dy_0\ e^{-i p_0 y_0} \overline{f}(x_0 + y_0, R(-\rho p_0) \vec{x} ),
    \label{invol-rho}
\end{align}
for any $f, g \in L^1(\mathbb{R}^3)$, where $R(\rho p_0)$ is a $2 \times 2$ rotation matrix with defining (dimensionless) parameter $\rho p_0$. The resulting associative $*$-algebra can then be extended to a suitable multiplier algebra of tempered distributions. Let $\mathcal{M}_\rho$ denotes this $*$-algebra. Useful relations involving the Lebesgue integral are $ \int d^4x\ (f \star_\rho g^\dag)(x)
    = \int d^4x\ f(x) \overline{g}(x)$, $ \langle f,g\rangle :=\int d^4x\ (f^\dag\star_\rho g)(x)=\int d^4x\ \overline{f}(x)g(x)$. Recall that $\int d^4x$ defines a trace as the group related to \eqref{coord-alg-intro} is $SE(2)$ which is unimodular.\\

A natural differential calculus can be obtained from the following set of (twisted) derivations of $\mathcal{M}_\rho$ \cite{gauge-rho}
\begin{equation}
    \mathfrak{D}=\big\{P_\mu:\mathcal{M}_\rho\to\mathcal{M}_\rho,\ \mu=0,3,\pm,\ \{ P_0,P_3 \}_{\bbone}\oplus \{ P_+\}_{\mathcal{E}_+}\oplus \{ P_- \}_{\mathcal{E}_-}\big\}\label{twisted-deriv},
\end{equation}
where the $P_\mu$'s act as the usual derivatives, i.e. $(P_\mu \triangleright f)(x) = -i \partial_\mu f(x),\ \mu=0,3,\pm$ and $P_\pm=P_1\pm iP_2$. They act as twisted derivations w.r.t the star-product $\star_\rho$, namely
\begin{eqnarray}
P_i(f\star_\rho g)&=&P_i(f)\star_\rho g+f\star_\rho P_i(g),\ i=0,3\nonumber\\
P_\pm(f\star_\rho g)&=&P_\pm(f)\star_\rho g+\mathcal{E}_\mp(f)\star_\rho P_\pm(g)\label{twist-leib}
\end{eqnarray}
for any $f,g\in\mathcal{M}_\rho$ with
\begin{equation}
 \mathcal{E}_\pm=e^{\pm i\rho P_0}   \label{lestwists}.
\end{equation}
The linear structures in \eqref{twisted-deriv} are defined from {\it{homogeneous}} linear combinations of elements of \eqref{twisted-deriv}, i.e. all the elements in the linear combination have the same twist degree defined 
by $ \tau(P_i)=0,\ i=0,3,\ \tau(P_\pm)=\pm1$. This gives rise to a grading extending to the differential calculus. Then, $\mathfrak{D}$ \eqref{twisted-deriv} inherits a structure of graded abelian Lie algebra and graded $\mathcal{Z}(\mathcal{{M}_\rho})$-bimodule so that one can write in obvious notations
\begin{equation}
    \mathfrak{D}=\mathfrak{D}_0 \oplus \mathfrak{D}_+\oplus\mathfrak{D}_-\label{grading},
\end{equation}
where $\mathfrak{D}_i$, $i=0,\pm$ can be read off from \eqref{twisted-deriv}. The construction of the resulting differential calculus can then be easily performed, see \cite{gauge-rho}.\\
Let $\mathbb{E}$ be a right (hermitian) module over $\mathcal{M}_\rho$. Define the 
twisted connection as a map $ \nabla:\mathfrak{D}_i\times\mathbb{E}\to\mathbb{E},\ i=0,\pm$ satisfying for any $m\in\mathbb{E}$, $f\in\mathcal{M}_\rho$
\begin{equation}
\nabla_{P_\mu+P^\prime_\mu}(m)=\nabla_{P_\mu}(m)+\nabla_{P^\prime_\mu}(m),\ \forall (P_\mu,P^\prime_\mu)\in\mathfrak{D}_i\times\mathfrak{D}_i, \ i=0,\pm\label{sigtaucon1}
\end{equation}
\begin{equation}
  \nabla_{z.P_\mu}(m)=\nabla_{P_\mu}(m)\star z,\ \forall P_\mu\in\mathfrak{D},\forall z\in Z(\mathcal{M }_\rho)\label{sigtaucon2},
\end{equation}
\begin{equation}
    \nabla_{P_\mu}(m\triangleleft f)=\nabla_{P_\mu}(m)\triangleleft f+{\beta}_{P_\mu}(m)\triangleleft P_\mu(f),\ \forall P_\mu\in\mathfrak{D}\label{sigtaucon3},
\end{equation}
where $m\triangleleft f$ denotes the action of the algebra on the module and the linearity condition \eqref{sigtaucon1} holds for linear combinations of derivations homogeneous in twist degree. As before, assume that 
$\mathbb{E}\simeq\mathcal{M}_\rho$ and the action of $\mathcal{M}_\rho$ on $\mathbb{E}\simeq\mathcal{M}_\rho$ is $m\triangleleft f=m\star f$. Finally, the map ${\beta}_{P_\mu}:\mathbb{E}\to\mathbb{E}$ in \eqref{sigtaucon3} can be determined by requiring that the following identity $\nabla_{P_\mu}((m\star_\rho f)\star_\rho g)=\nabla_{P_\mu}(m\star_\rho (f\star_\rho g))$ holds true, leading to 
\begin{equation}
    \beta_{P_\pm}=\mathcal{E}_\mp,\ \beta_{P_i}=\bbone,\ i=0,3\label{beta-fixe}.
\end{equation}
One arrives at
\begin{equation}
    \nabla_{P_i}(m\triangleleft f)=\nabla_{P_i}(m)\triangleleft f+P_i(f), i=0,3,\ \nabla_{P_\pm}(m\triangleleft f)=\nabla_{P_\pm}(m)\triangleleft f+\mathcal{E}_\mp(m)\triangleleft P_\pm(f),\ \label{sigtaucon4}
\end{equation}
for any $P_\mu\in\mathfrak{D}$. Set now $ A_\mu=\nabla_{P_\mu}(\bbone),\ \nabla_\mu:=\nabla_{P_\mu}$ as usual which finally gives rise to $\nabla_{\mu}(f)=A_{\mu}\star_\rho f+ P_\mu(f) $ for any $f\in\mathcal{M}_\rho$. By equipping the module with the canonical hermitean structure $h(m_1,m_2)=m_1^\dag\star m_2$ and requiring that twisted hermiticity conditions are verified, namely
\begin{equation}
(h(\mathcal{E}_+\triangleright\nabla_+(m_1),m_2)
    +h(\mathcal{E}_+\triangleright m_1,\nabla_+(m_2)))+ (+\to -)=P_+h(m_1,m_2)+P_-h(m_1,m_2)\label{hermiticity-cond1},
\end{equation}
\begin{equation}
    h(\nabla_i(m_1),m_2)
    +h(m_1,\nabla_i(m_2)))=P_ih(m_1,m_2),\ i=0,3\label{hermiticity-cond2}
\end{equation}
for any $m_1,m_2\in\mathbb{E}$, one obtains
\begin{equation}
    A_\pm^\dag=\mathcal{E}_\pm\triangleright A_\mp,\ A_i^\dag=A_i,\ i=0,3.\label{hermit-amu}
\end{equation}
The curvature $  \mathcal{F}(P_\mu,P_\nu):=\mathcal{F}_{\mu\nu}:\mathbb{E}\to\mathbb{E},\ \mu,\nu=0,3,\pm$ is
\begin{equation}
    \mathcal{F}_{\mu \nu} : = \mathcal{E}_\nu^{-1} \nabla_\mu \mathcal{E}_\nu \nabla_\nu - \mathcal{E}_\mu^{-1} \nabla_\nu \mathcal{E}_\mu \nabla_\mu,\ \ \mu,\nu=0,3,\pm\label{morph-curv}
\end{equation}
with $\mathcal{E}_\mu$ as in \eqref{beta-fixe} and satisfies $\mathcal{F}_{\mu\nu}=-\mathcal{F}_{\nu\mu},\ \mu,\nu=0,3,\pm$, and $\mathcal{F}_{\mu\nu}(m\star f)=\mathcal{F}_{\mu\nu}(m)\star f$. One arrives at
\begin{equation}
\mathcal{F}_{\mu\nu}(\bbone):=F_{\mu \nu} = P_\mu A_\nu - P_\nu A_\mu + (\mathcal{E}_\nu^{-1}\triangleright A_\mu) \star_\rho A_\nu - (\mathcal{E}_\mu^{-1}\triangleright A_\nu) \star_\rho A_\mu,\ \ \mu,\nu=0,3\pm\label{fmunu}.
\end{equation}
The group of unitary gauge transformations is again $\mathcal{U}=\{g\in\mathbb{E}\simeq\mathcal{M}_\rho,\ \ g^\dag\star g= g\star g^\dag= 1  \}$
and from the twisted gauge transformations for the connection
\begin{equation}
    \nabla^g_\mu(.)=(\mathcal{E}_\mu^{-1}\triangleright g^\dag)\star_\rho\nabla_\mu(g\star_\rho .)\label{gauge-connex},
\end{equation}
for any $g\in\mathcal{U}$, one obtains
\begin{equation}
A^g_\mu=(\mathcal{E}_\mu^{-1}\triangleright g^\dag)\star_\rho A_\mu\star_\rho g+(\mathcal{E}_\mu^{-1}\triangleright g^\dag)\star_\rho P_\mu g,\ \ F^g_{\mu\nu}=(\mathcal{E}_\mu^{-1}\mathcal{E}_\nu^{-1}\triangleright g^\dag)\star_\rho F_{\mu\nu} \star_\rho g \label{gaugetrans-amiou},
\end{equation}
where there is no summation over indices $\mu,\nu$ in the RHS of \eqref{gaugetrans-amiou}.\\
Anticipating the discussion of the next subsection, it is worth noticing that the expressions for the field strength $F_{\mu\nu}$ and gauge transformations will be formally similar to the $\kappa$-Minkowski case, upon changing $\mathcal{E}_\mu$ into the unique relevant twist for this latter case, i.e. $\mathcal{E}_\mu, \mu=0,3,\pm\to\mathcal{E}=e^{-(d-1)P_0}$.\\

The gauge invariant classical action is
\begin{equation}
    S_\rho:=\frac{1}{4g^2}\langle F_{\mu\nu}, F_{\mu\nu}\rangle =\frac{1}{4g^2}\int d^4x\  F_{\mu\nu}^\dag\star_\rho F_{\mu\nu}=\frac{1}{4g^2}\int d^4x\ \overline{F_{\mu\nu}}(x) F_{\mu\nu}(x)\label{action-class},
\end{equation}
where $g$ is a dimensionless coupling constant, $F_{\mu\nu}$ given by \eqref{fmunu},  $\overline{f}$ denotes the complex conjugate of $f$ and summation over $\mu, \nu$ is understood. \\
To show the invariance of \eqref{action-class} under the gauge transformations \eqref{gaugetrans-amiou}, compute $\langle F^g_{\mu\nu}, F^g_{\mu\nu}\rangle=\int d^4x\ F_{\mu\nu}^\dag\star_\rho\big[(\mathcal{E}_\mu^{-1}\mathcal{E}_\nu^{-1}\triangleright g^\dag)^\dag\star_\rho(\mathcal{E}_\mu^{-1}\mathcal{E}_\nu^{-1}\triangleright g^\dag)\big] \star_\rho F_{\mu\nu}$ where the cyclicity of the trace and $g\star_\rho g^\dag=\bbone$ have been already been used. Then, use $(t\triangleright f)^\dag=S(t)^\dag\triangleright f^\dag$ for any $t\in\mathcal{P}_\rho$, $f\in\mathcal{M}_\rho$ and $S(\mathcal{E}_\mu)=\mathcal{E}_\mu^{-1}$ (see eqn. \eqref{antipode}) to show that $(\mathcal{E}_\mu^{-1}\mathcal{E}_\nu^{-1}\triangleright g^\dag)^\dag=\mathcal{E}_\mu^{-1}\mathcal{E}_\nu^{-1}\triangleright g$ which combined with the second factor into the brackets gives $\mathcal{E}_\mu^{-1}\mathcal{E}_\nu^{-1}\triangleright(g\star_\rho g^\dag)=\bbone$ thanks to $g\star_\rho g^\dag=\bbone$.\\
The above gauge invariant 4-d action could be obviously constructed in principle in any $d\ge3$ dimensions by adding in the algebra of coordinates supplementary central coordinates. \\

One easily verifies \cite{gauge-rho} that the action \eqref{action-class} is invariant under the action of a deformed Poincar\'e algebra. In fact, a salient property is that $\mathcal{M}_\rho$ is a left-module algebra over a deformed Poincar\'e Hopf algebra $\mathcal{P}_\rho$ \cite{gauge-rho} for the action of $\mathcal{P}_\rho$ on $\mathcal{M}_\rho$, $\varphi:\mathcal{P}_\rho\otimes\mathcal{M}_\rho\to\mathcal{M}_\rho$, $\varphi(t\otimes f):=t\triangleright f$ given by $(P_\mu \triangleright f)(x) = -i \partial_\mu f(x)$, $(M_j \triangleright f)(x) = (\epsilon_{jk}^{l} x^{k}P_l \triangleright f )(x)$, $(N_j \triangleright f) = ( (x_0 P_j - x_j P_0) \triangleright f)(x)$.
In these latter relations, the indices for $(P_\mu, M_j,N_j)$ are such that $\mu \in \{0,+,-,3 \}$ and $j \in \{ +,-,3 \}$
where $M_j$ and $N_j$ are the rotations and boosts and the $P_\mu$'s are the translations, with $M_\pm=M_1+\pm iM_2$, $N_\pm=N_1\pm iN_2$, 
$P_\pm=P_1\pm iP_2$.\\
The deformed Poincar\'e Hopf algebra $\mathcal{P}_\rho$ is (the Lie algebra structure stays undeformed)
\begin{align}
\Delta(M_\pm) &= M_\pm \otimes \bbone + \mathcal{E}_\mp \otimes M_\pm,\ \Delta(N_\pm)= N_\pm \otimes \bbone + \mathcal{E}_\mp \otimes  N_\pm-\rho P_\pm\otimes M_3\nonumber\\
\Delta(M_3)&= M_3 \otimes \bbone + \bbone \otimes M_3,\ \Delta(N_3)=  N_3 \otimes \bbone + \bbone \otimes N_3-\rho P_3\otimes M_3\nonumber\\
\Delta(P_{0,3})&=P_{0,3}\otimes\bbone+\bbone\otimes P_{0,3},\  \Delta(P_\pm)=P_\pm\otimes\bbone+\mathcal{E}_\mp\otimes P_\pm,\ \Delta(\mathcal{E}_\pm)=\mathcal{E}_\pm\otimes\mathcal{E}_\pm \nonumber\\
\epsilon(P_\mu) &= 0,\  \quad \epsilon(\mathcal{E}) = 1,\ 
           \epsilon(M_j)= \epsilon(N_j) = 0,\ j=\pm,3,\nonumber\\
 S(P_0)&=-P_0,\ S(P_3)=-P_3,\ S(P_\pm) = - \mathcal{E}_\pm P_\pm,\  \ S(\mathcal{E}) = \mathcal{E}^{-1},\ S(M_3) = - M_3,\nonumber\\
           S(M_\pm)&=-\mathcal{E}_\pm M_\pm, \ S(N_3) = - N_3-\rho P_3M_3,\ S(N_\pm)=-\mathcal{E}_\pm N_\pm-\rho\mathcal{E}_\pm P_\pm M_3,\label{antipode}
           \end{align}
where $\Delta$, $\epsilon$ and $S$ are the coproduct, co-unit and antipode of $\mathcal{P}_\rho$. From this, one can check that any classical action $\int d^dx\ \mathcal{L}$, hence \eqref{action-class}, is invariant under the action of $\mathcal{P}_\rho$. Indeed, one has
\begin{equation}
    h\blacktriangleright\int d^4x\ \mathcal{L}:=\int d^4x\ h\triangleright \mathcal{L}=\epsilon(h)\int d^4x\ \mathcal{L},\label{decadixII}
\end{equation}
for any $h\in\mathcal{P}_\rho$, $\mathcal{L}\in\mathcal{M}_\rho$.\\

The one-loop properties of $S_\rho$ \eqref{action-class} can be explored upon BRST gauge-fixing. An unfortunate property is that the gauge-fixed action has a non-vanishing 1-loop tadpole (1-point function) for $A_\mu$ \cite{val-priv}, a pathology already encountered for almost all gauge theories on 
$\mathbb{R}^4_\theta$ and $\mathbb{R}^3_\lambda$. Again, the vacuum becomes unstable against quantum fluctuations.

\subsection{Yang-Mills type theory on $\kappa$-Minkowski space-time}\label{section42}

The popular $\kappa$-Minkowski space-time (in a d-dimensional case) \cite{luk2} can be viewed as generated by the Lie algebra of coordinates given by
\begin{equation}
[x_0,x_i]=\frac{\mathrm{i}}{\kappa}x_i, [x_i,x_j]=0,\ \ i,j=1,2,...,(d-1),
\end{equation}
where $\kappa>0$ has the dimension of a mass. The related group $\mathcal{G}$ is the affine group, $\mathcal{G}:=\mathbb{R}\ltimes\mathbb{R}^{d-1}$ which is non unimodular. The star product and involution obtained from subsection \ref{section21} are
\begin{align}
    (f\star_\kappa g)(x) &= \int \frac{dp^0}{2\pi} dy_0\ e^{-\mathrm{i}y_0p^0} f(x_0+y_0, \vec{x}) g(x_0, e^{-p^0/\kappa}\vec{x}), 
    \label{star-kappa}\\
    f^\dag(x) &= \int \frac{dp^0}{2\pi} dy_0\ e^{-\mathrm{i}y_0p^0} \overline{f} (x_0+y_0, e^{-p^0/\kappa} \vec{x})
    \label{invol-kappa},
\end{align}
for any $f\in\mathcal{M}_\kappa$, the $*$-algebra modeling the $\kappa$-Minkowski space-time. Useful relations for the Lebesgue integral are $\int d^dx\ (f\star_\kappa g^\dag)(x)=\int d^dx\ f(x){\bar{g}}(x)$, $\langle f,g\rangle:=\int d^dx\left(f^\dag\star_\kappa g\right)(x)$. Besides, one has
\begin{equation}
    \int d^dx\ (f\star_\kappa g)(x) = \int d^dx\ ((\mathcal{E}^{d-1}\triangleright g) \star_\kappa  f)(x),\ \ 
    \label{twisted-trace}
\end{equation}
where
\begin{equation}
\mathcal{E}=e^{-P_0/\kappa}\label{cestE}
\end{equation}
so that the Lebesgue integral is no longer a trace but now defines a KMS weight \cite{kuster-a}, \cite{kuster-b} for the modular group of automorphisms $\sigma_t(f)=e^{it(d-1)P_0/\kappa}\triangleright f$. For a discussion on the link with the Tomita-Takesaki modular theory and the possibility to define a global time observer independent see \cite{conn-rov}, \cite{PW2018}, \cite{khersent}. \\
A twisted differential calculus can be easily obtained from the Lie algebra of twisted derivations \cite{MW2020a}
\begin{equation}
    \mathfrak{D}=\big\{X_\mu:\mathcal{M}_\kappa\to \mathcal{M}_\kappa,\ \  X_0=\kappa(1-\mathcal{E}),\ \ X_i=P_i,\ \  i=1,2,...,d-1\big\}
\label{tausig-famil},
\end{equation}
which is also a $\mathcal{Z}(\mathcal{M}_\kappa)$-bimodule. The derivations satisfy
\begin{equation}
    X_\mu(f\star_\kappa h) = X_\mu(f) \star_\kappa h +  \mathcal{E}(f) \star_\kappa X_\mu(h),
    \label{tausigleibniz}
\end{equation}
for any $f,h\in\mathcal{M}_\kappa$. Note that the $X_\mu$'s are not real derivations, as $(X_\mu(f))^\dag = -\mathcal{E}^{-1} (X_\mu(f^\dag))$.\\
A twisted connection, curvature and gauge transformations can then be defined from a straightforward adaptation of the construction performed in subsection \ref{section41}, of course under the same assumptions as in subsection \ref{section41}. In particular, the curvature $F_{\mu\nu}$ is still given by \eqref{fmunu} in which one sets $\mathcal{E}_\mu=\mathcal{E}$, where $\mathcal{E}$ given by \eqref{cestE}, while the gauge group is still $\mathcal{U}=\{g\in\mathbb{E}\simeq\mathcal{M}_\kappa,\ \ g^\dag\star_\kappa g= g\star_\kappa g^\dag= 1  \}$. The gauge transformations of the curvature are
\begin{equation}
F_{\mu\nu}^g = \mathcal{E}^2(g^\dag) \star_\kappa F_{\mu\nu} \star_\kappa  g.
\end{equation}
Now, the following classical action 
\begin{equation}
   S_\kappa(F_{\mu\nu})=\int d^dx\ F_{\mu\nu}\star_\kappa  F_{\mu\nu}^\dag,\label{action-kappa}
\end{equation}
is gauge invariant provided
\begin{equation}
    \mathcal{E}^{d-1-2}(g)\star_\kappa \mathcal{E}^2(g^\dag)=\bbone,
\end{equation}
which holds true only when $d=5$ \cite{MW2020b}. Notice that \eqref{action-kappa} is also invariant under the well-known $\kappa$-deformation of the Poincar\'e algebra \cite{MR1994}; indeed, a relation similar to \eqref{decadixII} holds with however $\mathcal{P}_\rho$ replaced by the $\kappa$-Poincar\'e Hopf algebra, \cite{DS}, \cite{MW2020a}.\\
One concludes that within the present framework, a quadratic action in the field strength is gauge invariant only in five dimensions. Hence, a physically salient prediction for this gauge theory on $\kappa$-Minkowski space-time is the existence of one extra-dimension. This property comes from the loss of cyclicity of the trace, \eqref{twisted-trace}, and the fact that the twist appearing in \eqref{twisted-trace} depends on the (engineering) dimension of the $\kappa$-Minkowski space-time. It is worth pointing out that a similar behavior can already be expected for the non unimodular cases collected in the Table 1: in these quantum space-times, gauge invariance of actions of Yang-Mills type obtained from a construction {\it{similar}} to the one presented in the present discussion could not be reached in four dimensions.\\
Upon gauge-fixing \cite{MW2021}, some one-loop properties of the present gauge theory can be explored. Unfortunately, the resulting gauge-fixed action has a non-vanishing 1-loop tadpole (1-point function) for $A_\mu$ \cite{HMW2022b}, so that again the vacuum is unstable. Note that attempt to balance the pure gauge contributions to the tadpole by coupling matter to the theory have been unsuccessful in this gauge theory.\\

To conclude this subsection, it is worth pointing out that the above $d=5$ dimensional prediction is {\it{not}} a universal prediction characterizing any gauge theory on $\kappa$-Minkowski space. Indeed, this interesting result depends obviously of the assumptions underlying the construction, in particular of the choice of the differential calculus, the assumed general form of the action as the square of a curvature and by the way the assumed description of the noncommutative connection.

\section{Discussion}\label{section5}

Let me summarize the present situation. The noncommutative gauge theories considered in this note can be viewed as generalizations of Yang-Mills theory, even in some sense in their matrix model formulation (if any) at least as they are all based on a noncommutative analog of a usual gauge connection. The almost common feature is that severe unsolved problems remain unsolved already at the one-loop level when trying to go beyond the classical order. \\

In the case of Moyal space $\mathbb{R}^4_\theta$, no all order perturbatively renormalisable gauge theory has been obtained so far, due to a resistant UV/IR mixing and possibly supplemented by a non zero one-loop tadpole for the gauge potential implying an unstable vacuum. Note however that a recent approach based on braided $L_\infty$-algebra \cite{marija24}, implying however a complete change of formalism, succeeded in eliminating the mixing and fulfilling the Slavnov-Taylor identities at one-loop. This possible way out may be an interesting alternative and is worth exploring.\\ The situation is better for gauge theories on $\mathbb{R}^3_\lambda$ but unfortunately limited to a 3-dimensional case (and in some instance non standard commutative limits). One family of all orders finite gauge matrix models has been exhibited which in some sense borrow some features of the LSZ model, one gauge model of this family being solvable. It may eventually be interesting to determine whether or nor some of these models may be related to some Group Field Theory description of the 3-dimensional (Loop) Quantum Gravity, which is not clear. Note that Yang-Mills generalizations on $\mathbb{R}^3_\lambda$ have non vanishing 1-loop tadpole.\\
Yang-Mills type models have been built on only two representatives among the 11 new quantum Minkowski space-times recently characterized through tractable star-products and involutions and again suffer from non-vanishing tadpoles at one-loop. It is not know if some UV/IR mixing shows up but a non zero tadpole already makes their interest questionable, unless a way to cope with is found. What about Yang-Mills models on the other quantum space-times? Some of these quantum space-times do not support a trace but a KMS weight instead, sometimes called a "twisted trace", as $\kappa$-Minkowski, which combined with gauge-invariance requirement of the action put a strong constraint on the number of dimensions. It can be already expected that $d=4$ is excluded for these quantum space-times. The physical consequences of the appearance of a KMS weight as constituent ingredient of the gauge invariant actions have been poorly explored in these models and should be examined, e.g. the possibility for a global time observer independent to appear. \\

To go beyond these noncommutative Yang-Mills models, the present formalism can be modified in order to accommodate at least a noncommutative analog of a linear connection. This has been achieved recently and will be reported in a forthcomming publication \cite{HW34}.

\section*{Acknowledgments}
I am very grateful to the organizers of the conference "Applications of Noncommutative Geometry to Gauge Theories, Field Theories, and Quantum Space-Time", CIRM (2025) for their invitation. \\
Discussions and exchanges during past collaborations and meetings with P. Bieliavsky, M. Dubois-Violette, K. Hersent, T. Juri\'c, F. Lizzi, V. Maris, P. Martinetti, T. Masson, Ph. Mathieu, F. Po\v{z}ar, A. Sitarz, D. Vassilevich, P. Vitale, A. Wallet are gratefully acknowledged. \\
I also thank the Actions CA21109 CaLISTA ``Cartan geometry, Lie, Integrable Systems, quantum group Theories for Applications'' and CA23130 BridgeQG ''Bridging high and low energies in search of quantum gravity" from the European Cooperation in Science and Technology.

\begin{table}
        \makebox[\textwidth][l]{ 
        \vspace{-1cm}
        \hspace{-2.5cm}
        \pgfplotstabletypeset[normal]{ %
         Case & Coordinates & $x^M$ &Commutators &  Group  & Matrix $A(p^M)$   \\
        & & & \textcolor{blue}{\textsc{Unimodular groups}} & & \\[-1ex]
        10 &\pbox{5cm}{$x^0 = z-t,\; x^1 = x$,\\$x^2 = z-t-y,$\\$x^3 = z.$} &$x^M = x^1$& \pbox{5cm}{$[x^M,x^2]=-i\lambda x^0$ }& $\mathbb{H} $ & \pbox{4cm}{\tiny$\left(\begin{matrix}1 & 0 & 0 & 0\\0 & 1 & 0 & 0\\\lambda p^{1} & 0 & 1 & 0\\0 & 0 & 0 & 1\end{matrix}\right) $\normalsize} \\[3ex]
        \hline&&&&& \\[-4ex]
        11 & \pbox{5cm}{$x^0 = z-t$,\\$x^1 = x$,\;$x^2 = y$,\\ $x^3 = -z.$}&$x^M = x^2$&\pbox{5cm}{$[x^M,x^1]=-i\lambda x^0$\\
        $[x^M,x^3]= -i\lambda x^1$} & $\mathcal{G}_{4,1}$ & \pbox{4cm}{\tiny$\left(\begin{matrix}1 & 0 & 0 & 0\\\lambda p^{2} & 1 & 0 & 0\\0 & 0 & 1 & 0\\\frac{\lambda^{2} \left(p^{2}\right)^{2}}{2} & \lambda p^{2} & 0 & 1\end{matrix}\right)\normalsize
$} \\[4ex]
        \hline&&&&& \\[-4ex]
        12 & \pbox{5cm}{$x^0 = t$,\\$x^1 = x$,\;$x^2 = y$,\\ $x^3 = t-z.$} &$x^M = x^1$&\pbox{5cm}{$[x^M,x^0]=-i\lambda x^3 $ }& $\mathbb{H} $ & \pbox{4cm}{\tiny$\left(\begin{matrix}1 & 0 & 0 & \lambda p^1\\0 & 1 & 0 & 0\\0 & 0 & 1 & 0\\0 & 0 & 0 & 1\end{matrix}\right) $\normalsize} \\[3ex]
        \hline&&&&& \\[-4ex]
        13& \pbox{5cm}{$x^0 = t,\; x^1 = x$,\\$x^2 = y,\;x^3 = z.$} &$x^M = x^0$&\pbox{5cm}{$[x^M,x^1]=i \lambda x^2$ \\$ [x^M,x^2]=-i \lambda x^1$} &$SE(2) $ & \pbox{4cm}{\tiny$\left(\begin{matrix}1 & 0 & 0 & 0\\0 & \cos{\left(\lambda p^{0} \right)} & - \sin{\left(\lambda p^{0} \right)} & 0\\0 & \sin{\left(\lambda p^{0} \right)} & \cos{\left(\lambda p^{0} \right)} & 0\\0 & 0 & 0 & 1\end{matrix}\right)$\normalsize} \\[3ex]
        \hline&&&&& \\[-4ex]
        14 & \pbox{5cm}{$x^0 = t,\; x^1 = x$,\\$x^2 = y,\;x^3 = z.$} &$x^M = x^3$&\pbox{5cm}{$[x^M,x^1]=i \lambda x^2$ \\$ [x^M,x^2]=-i \lambda x^1$} &$SE(2) $ &\pbox{4cm}{\tiny$\left(\begin{matrix}1 & 0 & 0 & 0\\0 & \cos{\left(\lambda p^{3} \right)} & - \sin{\left(\lambda p^{3} \right)} & 0\\0 & \sin{\left(\lambda p^{3} \right)} & \cos{\left(\lambda p^{3} \right)} & 0\\0 & 0 & 0 & 1\end{matrix}\right)$\normalsize}\\[3ex]
        \hline&&&&& \\[-4ex]
        15 & \pbox{5cm}{$x^0 = t+z$,\\$x^1 = x$,\;$x^2 = y$,\\ $x^3 = z.$} &$x^M = x^0$&\pbox{5cm}{$[x^M,x^1]=i \lambda x^2$ \\$ [x^M,x^2]=-i \lambda x^1$} &$SE(2) $ & \pbox{4cm}{\tiny$\left(\begin{matrix}1 & 0 & 0 & 0\\0 & \cos{\left(\lambda p^{0} \right)} & - \sin{\left(\lambda p^{0} \right)} & 0\\0 & \sin{\left(\lambda p^{0} \right)} & \cos{\left(\lambda p^{0} \right)} & 0\\0 & 0 & 0 & 1\end{matrix}\right)$\normalsize}\\[3ex]
        \hline&&&&& \\[-4ex]
        16 & \pbox{5cm}{$x^0 = t,\; x^1 = x$,\\$x^2 = y,\;x^3 = z.$} &$x^M = x^1$&\pbox{5cm}{$[x^M,x^0]=i\lambda x^3 $\\$ [x^M,x^3]=i\lambda x^0$ }& $SE(1,1) $ & \pbox{5cm}{\tiny$\left(\begin{matrix}\cosh{\left(\lambda p^{1} \right)} & 0&0&-\sinh{\left(\lambda p^{1} \right)} \\0 & 1 &0 & 0 \\0 & 0 & 1 & 0\\ -\sinh{\left(\lambda p^{1} \right)} & 0 & 0 & \cosh{\left(\lambda p^{1} \right)}\end{matrix}\right)$\normalsize} \\[4ex]
        \hline\hline & & & \textcolor{blue}{\textsc{Nonunimodular groups}} & & \\
        \pbox{5cm}{7 \\$(\alpha = \frac{1}{\zeta})$} &\pbox{5cm}{$x^0 = \alpha t$,\\$x^1 = x,\; x^2 = y$,\\ $x^3 = \alpha(z - t)$} &$x^M = x^0$& \pbox{5cm}{$[x^M,x^1]=i\lambda(\alpha x^1 + x^2),$\\$[x^M,x^2]=i\lambda(\alpha x^2-x^1)$\\ $[x^M,x^3]=i\lambda\alpha x^3$} & $\mathcal{G}_{4,6}^{\alpha, \alpha}$ & \hspace{-10mm} \pbox{4cm}{\tiny$\left(\begin{matrix}1 & 0& 0 & 0\\0 & e^{- \alpha \lambda p^{0}} & 0 & 0\\0 & 0 & e^{- \alpha \lambda p^{0}} \cos(\lambda p^{0}) & -e^{- \alpha \lambda p^{0}} \sin(\lambda p^{0})\\0 & 0 & e^{- \alpha \lambda p^{0}} \sin(\lambda p^{0}) & e^{- \alpha \lambda p^{0}} \cos(\lambda p^{0})\end{matrix}\right)$\normalsize}\\[4ex]
        \hline&&&&& \\[-3ex]
        8& \pbox{5cm}{$x^0 = t$,\\$x^1 = x,\; x^2 = y$,\\ $x^3 = \zeta(t - z)$} &$x^M = x^0$&\pbox{8cm}{$[x^M,x^1]=i\lambda(x^3 + x^1)$\\$ [x^M,x^2]=i\lambda x^2$\\$ [x^M,x^3]=i\lambda x^3$}& $\mathcal{G}_{4,2}^{1}$ &  \pbox{4cm}{\tiny$\left(\begin{matrix}1 & 0 & 0 & 0\\0 & e^{- \lambda p^{0}} & 0 & - \lambda p^{0} e^{- \lambda p^{0}}\\0 & 0 & e^{- \lambda p^{0}} & 0\\0 & 0 & 0 & e^{- \lambda p^{0}}\end{matrix}\right)$\normalsize}\\[4ex]
        \hline&&&&& \\[-4ex]
        17 & \pbox{5cm}{$x^0 = -t$,\\$x^1 = x,\;x^2 = y$,\\ $x^3 = z-t$} &$x^M = x^0$&$[x^M,x^3] = i\lambda x^3$& $\mathcal{G}_{2,1}$ & \pbox{4cm}{\tiny$\left(\begin{matrix}1 & 0 & 0 & 0\\0 & 1 & 0 & 0\\0 & 0 & 1 & 0\\0 & 0 & 0 & e^{- \lambda p^{0}}\end{matrix}\right)$\normalsize}\\[3ex]
        \hline&&&&& \\[-4ex]
        \pbox{5cm}{18 \\$\alpha = \frac{1}{\zeta}$} &\pbox{5cm}{$x^0 =\alpha t$,\\$x^1 = x-y$,\\$x^2 = x+y$,\\ $x^3 =t- z.$} & $x^M = x^0$&\pbox{8cm}{$[x^M,x^1]=i\lambda x^2$\\$ [x^M,x^2]=-i\lambda x^1$\\$ [x^M,x^3]=i\lambda x^3$}& $\mathcal{G}_{4,6}^{1, 0}$ &   \pbox{4cm}{\tiny$\left(\begin{matrix}1 & 0 & 0 & 0\\0 & \cos{\left(\lambda p^{0} \right)} & - \sin{\left(\lambda p^{0} \right)} & 0\\0 & \sin{\left(\lambda p^{0} \right)} & \cos{\left(\lambda p^{0} \right)} & 0\\0 & 0 & 0 & e^{- \lambda p^{0}}\end{matrix}\right)$\normalsize}\\
}}
\caption{\textbf{Quantum Minkowski space-times} The numbers in the leftmost column corresponds to the space-times numbers of \cite{zak}, the next column refers to the coordinate change to map the bracket of \cite{mercati-a}, \cite{mercati-b} to the \eqref{general-liecommut} form of the Lie algebra. The commutation relations and associated Lie groups are listed in the next two columns. The rightmost column gives the $A(p^M)$ matrices, with parameter $p^M$ encoding the time/light/space-like noncommutativity.}
\end{table}
\vfill\eject

\end{document}